\documentstyle[12pt, fleqn]{article}
\def\ownref{\par\noindent\hangindent=6mm\hangafter=1}

\begin{document}
\baselineskip 8mm

\begin{center}

{\bf Wavelet Space-Scale-Decomposition Analysis of QSO's Ly$\alpha$

Absorption Lines: Spectrum of non-Gaussianity}

\bigskip
\bigskip
\bigskip

Jesus Pando and Li-Zhi Fang

\bigskip

Department of Physics, University of Arizona, Tucson, AZ 85721

\end{center}

\newpage

\begin{center}

{\bf Abstract}

\end{center}

Using a discrete wavelet based space-scale decomposition (SSD), the spectrum
of the skewness and kurtosis is developed to describe the non-Gaussian
signatures in cosmologically interesting samples.
Because the basis of the discrete wavelet is compactly supported,
the one-point distribution of the father function coefficients (FFCs) taken
from one realization is a good estimate of the probability
distribution function (PDF) of density fields if the ``fair sample
hypothesis" holds. These FFC one-point distributions can also avoid the
constraints of the central limit theorem on the detection of
non-Gaussianity. Thus, the FFC one-point distributions are effective in
detecting
non-Gaussian behavior in samples such as non-Gaussian clumps embedded in a
Gaussian background, regardless of the number or number
density of the clumps. We demonstrate that the non-Gaussianity spectrum can
reveal not only the magnitudes, but also the scales of the non-Gaussianity.

Also calculated are the FFC one-point distributions, skewness and kurtosis
spectra for real data and linearly simulated samples of QSO Ly$\alpha$
forests. When considering only second and lower order of statistics,
such as the number density and two-point correlation functions, the
simulated data
show the same features as the real data. However, the kurtosis
spectra of samples given by different models are found to be different.
On the other hand, the spectra of skewness
and kurtosis for independent observational data sets are found to be the
same. Moreover, the real data are significantly different from the
non-Gaussianity
spectrum of various possible random samples. Therefore, the non-Gaussian
spectrum
is necessary and valuable for model discrimination.


\noindent{\it Subject headings:} cosmology: large-scale structure of
Universe  -- quasars: absorption lines

\newpage

\noindent{\bf 1. Introduction}

\bigskip

This is our third paper on developing the discrete wavelet transform
(DWT) into a powerful space-scale decomposition (SSD) method for the analysis
of the large scale structures. The first two papers studied 1) identification
of clusters (Pando \& Fang 1996, hereafter PF1); 2) determination of power
spectrum of density perturbations (Pando \& Fang 1995, hereafter PF2). In
this paper, a method of detecting non-Gaussian behavior via the DWT, and
its application to both simulated and observational data of QSO's
Ly$\alpha$ absorption lines is presented.

There are important motivations for studying the non-Gaussianity of cosmic
distributions.
In standard inflation/dark matter cosmology, primordial perturbations
generated in the inflationary era are scale-free and Gaussian. The subsequent
evolution of the density perturbations in the linear regime destroys the
scaling of the spectrum, but the density field remains Gaussian. Deviations
from
the Gaussian state in the cosmic density distribution occur first because
of the non-linear evolution caused by the gravitational instability of
baryonic
and dark matter. In this highly non-linear evolutionary stage the density
field
should be very non-Gaussian. Additionally, even when the background density
distribution is Gaussian, the distribution of visible objects may  be
non-Gaussian if their distribution is biased with respect to the background
distribution. However, detection of non-Gaussian
behavior has, in many respects, been inconclusive.  For instance,
no deviations
from Gaussian behavior were found in the QDOT-IRAS redshift survey
(Feldman, Kaiser \&
Peacock 1994), and even for simulation samples in which a strong non-linear
evolution is part of the model, the one point distribution function was
found to be consistent with a Gaussian distribution (Suginohara \& Suto
1991).

It is generally believed that the failure in detecting non-Gaussian
behavior is due partially to the constraint imposed on Fourier techniques by
the central limit theorem when applied to the one point distribution
function.
It is well known from the central limit
theorem of random fields (Adler 1981; Ivanonv \& Leonenko 1989) that if
the universe consists of a large number of dense clumps, and those clumps
are independent, the real and imaginary components of each individual Fourier
mode are Gaussian distributed although the probability distribution
functions (PDF) of the field itself are highly non-Gaussian. If the
clumps are
not distributed
independently, but are correlated, the central limit theorem still holds
if the two-point correlation function of the clumps approaches zero
sufficiently fast (Fan \& Bardeen 1995). The central limit theorem
holds even for processes that are a time (or space) average of a non-Gaussian
distribution as long as the ratio of the sampling time to the scale of the
fluctuation increases (Adler 1981). For these reasons, the one-point
distribution
function of Fourier modes is not a sensitive enough measure
to detect deviations from Gaussian behavior.

This difficulty can be overcome by using the count in cell (CIC) method
because
the CIC is based on localized window functions, which in essence, keep
the sampling time to scale of fluctuation ratio mentioned above from
increasing. Thus, the CIC statistic is not restricted by the central limit
theorem. CIC has succeeded in detecting non-Gaussian signatures
(Hamilton 1985; Alimi, Blanchard \& Schaeffer 1990; Gazta\~naga \& Yokoyama
1993; Bouchet et al. 1993; Kofman et al. 1994; Gazta\~naga \& Frieman 1994).
However, it has been found that the CIC results are dependent on the
parameters of the window function (Juszkiewicz et al. 1993). This is
because non-Gaussian distributions are generally scale-dependent, and
therefore, a window function with a different scale will obtain different
degrees of non-Gaussian behavior (Yamada \& Ohkitani 1991). The results
obtained via a CIC analysis will be a superposition of information on
scales larger than the size of the window. Therefore, in order to
completely describe the non-Gaussianity of density fields and object
distributions, it is very necessary to have an effective and uniform
measure of the scale-dependence or {\em spectrum} of non-Gaussianity.

In the first part of this paper, we will study a method of detecting
the non-Gaussian spectrum by the discrete wavelet SSD. As with the CIC,
the bases of discrete wavelet SSD
are localized, and therefore, avoid the restrictions of
central limit theorem. On the other hand, the SSD modes are orthogonal
and complete and no scale mixing occurs. It is
easy to decompose the contribution of structures on various scales to the
non-Gaussianity.
Hence, one can expect that a discrete wavelet SSD will be  effective in
detecting the non-Gaussianity spectrum of large scale structures.

In the second part of the paper, we calculate the non-Gaussianity of real
and simulated samples of QSO Ly$\alpha$ forests. Most 2nd order statistical
studies, such as the two-point correlation function or power spectrum, have
failed to detect structure in the distributions of QSO Ly$\alpha$ forests. On
the other hand, statistics not based on
the power spectrum or 2 point correlation function have indicated
that these distributions should have structure  (see, for example,
Duncan, Ostriker, \& Bajtlik 1989; Liu and Jones 1990;  Fang 1991).
These examples illustrate that it is necessary to go beyond the 2nd order
statistics, that is, to look at non-Gaussian behavior.

\bigskip

\noindent{\bf 2. Method}

\bigskip

\noindent{\it 2.1 Wavelet expansion}

\bigskip

We use the notation introduced in PF1 and PF2. Consider a
one-dimensional density  field
$\rho(x)$ over a range $0 \leq x \leq L$. It is convenient to use the density
contrast defined by
\begin{equation}
\delta(x) = \frac{\rho(x) - \bar\rho}{\bar\rho}
\end{equation}
where $\bar\rho$ is the mean density in this field. To express $\delta(x)$
in a Fourier expansion, we take the convention
\begin{equation}
\delta(x)=\sum_{n = - \infty}^{\infty} \delta_n e^{i2\pi nx/L}
\end{equation}
with the coefficients computed by
\begin{equation}
\delta_n= \frac{1}{L}\int_0^{L} \delta(x)e^{-i2\pi nx/L}dx.
\end{equation}

Now contrast this with the wavelet expansion of the
density $\rho(x)$ (Meyer 1993). We first assume that $\rho(x)$ is an $L$
periodic function defined on
space $- \infty < x < \infty$ (this condition may actually be relaxed.).
The wavelet expansion is then given by (Daubechies 1992, PF2)
\begin{equation}
\rho(x) = \sum_{m=-\infty}^{\infty} c_{0,m}\phi_{0,m}(x) +
\bar{\rho} \sum_{j=0}^{\infty} \sum_{l= - \infty}^{\infty}
 \tilde{\delta}_{j,l} \psi_{j,l}(x)
\end{equation}
where $\bar\rho$ is the mean density. $c_{0,m}$ is the mother function
coefficient (MFC) on scale $j=0$ and position $l$, while
$\tilde{\delta}_{j,l}$
is the father function coefficient (FFC) on scale $j$ and position $l$.
The MFC and FFC are calculated by the inner products as
\begin{equation}
c_{j,m}=\int_{-\infty}^{\infty} \rho(x) \phi_{j,m}(x) dx
\end{equation}
\begin{equation}
\tilde{\delta}_{j,l}=\int_{-\infty}^{\infty} \delta(x) \psi_{j,l}(x)dx
\end{equation}
where $\delta(x)$ is the density contrast given by eq.(1).
The mother function $\phi_{j,m}(x)$ and father function $\psi_{j,l}$
are given, respectively, by $\phi_{j,m}(x) =(2/L)^{1/2} \phi(2^jx/L - m)$ and
$\psi_{j,l}(x) =(2^j/L)^{1/2} \psi(2^jx/L-l)$, where $\phi(x)$ is the
scaling function, and $\psi(x)$ the basic wavelet. The
$\phi(x)$ and $\psi(x)$ must meet certain admissibility
conditions (Farge 1992).  The Daubechies 4 wavelets
meet all these conditions and are the discrete wavelet bases used in this
paper.

\bigskip

\noindent{\it 2.2 Restriction of central limit theorem}

\bigskip

Among the various methods of detecting Gaussian deviations, the one-point
distribution of the density field is especially important because the
probability distribution functions (PDF) of density fields can be directly
determined by the one-point distribution. That is, the one-point distribution
can detect not only the deviation from a Gaussian distribution, but can
also detect the non-Gaussian PDF itself.

Let us consider the non-Gaussianity of density fields
$\delta(x)$ consisting of randomly distributed non-Gaussian clumps.
In this case, eq.(3) shows that for large $L$ the Fourier amplitudes,
$\delta_n$, are given by a superposition of a large number of non-Gaussian
distributions. According to the central limit theorem, the distribution
of $\delta_n$ will be Gaussian when the total number of clumps
is large. Thus, in general, the non-Gaussianity of distributions of randomly
distributed clumps can not be seen from the one-point
distribution of the Fourier modes, $\delta_n$, even if the PDF
function of clumps is highly non-Gaussian.

On the other hand, the father functions, $\psi_{j,l}(x)$, are localized.
If the scale of the clump is $d$, eq.(6) shows that the FFC,
$\tilde{\delta}_{j,l}$, with $j=\log_2(L/d)$,  is determined only by
the density field in a range containing no more than one clump.
That is, FFCs are not given by a superposition of a large number of
non-Gaussian
variables, but determined by one, or at most, a few non-Gaussian processes.
The one point distribution of the FFC, $\tilde{\delta}_{j,l}$,
avoids the restriction of the central limit theorem, and is able to detect
non-Gaussianity, regardless the total number of the clumps in the sample
being considered.

One can study this problem from the orthonormal basis used for the
expansion of the density field. A key condition needed for the
central limit theorem to hold is that the modulus of the basis be less
than $C/\sqrt L$, where $L$ is the size of the sample and $C$ is a constant
(Ivanonv \& Leonenko 1989). Obviously, all Fourier-related orthonormal
bases satisfy this condition because the Fourier orthonormal bases
in 1-dimension are such that $(1/\sqrt L) |\sin kx| < C/\sqrt L$
and $(1/\sqrt L)|\cos kx| < C/\sqrt L$ and $C$ is independent of
coordinates in both physical space $x$ and scale space $k$. On the other
hand, the father functions (5) have
\begin{equation}
|\psi_{j,l}(x)| \sim \left( \frac{ 2^{j}}{L}\right )^{1/2} {\rm O}(1)
\end{equation}
because the magnitude of the basic wavelet $\psi(x)$ is of the order
1. The condition $|\psi_{j,l}(x)| < C/ \sqrt{L}$, will no longer hold
for a constant $C$ independent of scale variable $j$.

Aside from the Daubechies 2 wavelet (i.e.,the Haar wavelet),
the father functions for the Daubechies wavelets are also
localized in Fourier space. The FFC $\tilde{\delta}_{j,l}$ is only
determined
by the perturbations on scale $j$, regardless of perturbations on other
scales. If the universe consists of non-Gaussian clumps on various
scales $j$, the FFCs $\tilde{\delta}_{j,l}$ will still be effective in
detecting the non-Gaussian signal. However, a measure given by a sum over
a large number of scales will fail to do so.

The mother functions of the discrete wavelet transform are localized in
spatial space. But the mother functions $\phi_{j,l}(x)$ are not orthogonal
with respect to the scale index $j$, i.e. they are  not localized in Fourier
space (PF1, PF2). This means that the MFCs
are given by a sum over perturbations on all scales larger than $L/2^j$.
Thus, if the number of the independent
clumps in Fourier (scale) space is large, the MFCs will also be
Gaussian by the central limit theorem. The count in cell (CIC) analysis
is essentially the same as the MFC. The window function of the CIC
corresponds to the mother function, and the count to the amplitude of
the MFCs.  Like the MFCs, the CIC is scale-mixed and it may not always be
sensitive to the non-Gaussian behavior on specific scales.

\bigskip

\noindent{\it 2.3 One-point distribution of FFCs}

\bigskip

For the density field to be consistent with the cosmological
principle, $\delta(x)$
should be described by a homogeneous random process, i.e. its probability
distribution function should remain the same when $x$ translated.
On the other hand, the bases functions $\psi_{j,l}(x)$ are orthogonal
to translation $l$. Equation (6) shows that
the probability distribution of $\tilde{\delta}_{j,l}$ should be
$l$-independent.

The $l$-independence can also be seen from the relationship between
FFC $\tilde{\delta}_{j,l}$ and Fourier coefficients $\delta_n$ (PF2)
given as
\begin{equation}
\tilde{\delta}_{j,l} = \sum_{n = - \infty}^{\infty}
\left(\frac{ 2^{j}}{L}\right )^{-1/2} \delta_n
\hat{\psi}(-n/2^j)  e^{i2\pi nl/2^j}
\end{equation}
where $\hat{\psi}$ is the Fourier transform of $\psi$. The
wavelet basis functions $\psi_{j,l}(x)$ are compactly supported
in both $x$ and Fourier space. Generally, $\hat{\psi}(n)$ will have two
symmetric peaks with centers at $n= \pm n_p$, and with width $\Delta n_p$.
The sum over integer $n$ in eq.(8) need only be taken on two
ranges of
$(n_p - 0.5 \Delta n_p)2^j \leq n \leq (n_p + 0.5\Delta n_p)2^j$ and
$-(n_p + 0.5 \Delta n_p )2^j \leq n \leq -(n_p - 0.5 \Delta n_p )2^j$.
Eq.(8) can be approximately rewritten as
\begin{eqnarray*}
\tilde{\delta}_{j,l} & \simeq & \left(\frac{L}{2^{j}} \right)^{1/2}
  2\sum_{n=(n_p - 0.5\Delta n_p)2^j}^{(n_p + 0.5\Delta n_p)2^j}
  {\rm Re}\{ \hat{\psi}(n_p)\delta_{n} e^{i2\pi nl/2^j} \}
\end{eqnarray*}
 \begin{equation}
\ \ \ \ \ \ \ \ \ \ \simeq \left(\frac{L}{2^{j}}\right)^{1/2}
|\hat{\psi}(n_p)|
2\sum_{n=(n_p - 0.5\Delta n_p)2^j}^{(n_p + 0.5\Delta n_p)2^j}
 |\delta_{n}|\cos (\theta_{\psi}+\theta_{n}+2\pi nl/2^j)
\end{equation}
where we have used $\hat{\psi}(-n_p)=\hat{\psi}^*(n_p)$ and
$\delta_{-n}= \delta^*_{n}$, because both $\psi(x)$ and $\delta(x)$ are
real. $\theta_{\psi}$, $\theta_n$ in eq.(9) are the phases of
$\hat{\psi}(n_p)$ and $ \delta_n$, respectively. As we pointed out
in last section, $\delta_n$ is Gaussian even when the clumps are
non-Gaussian. The phase of $\delta_n$, $\theta_n$, should be
uniformly randomly distributed, and therefore, from eq.(9), the
probability  distribution of $\tilde{\delta}_{j,l}$ is independent of $l$.

Moreover, if the spatial correlations of the random field $\delta(x)$
decay sufficiently rapidly with increasing separation, ranges
with different $l$ are essentially statistically independent. That is,
even for one
realization of $\delta(x)$, the values of FFCs $\tilde{\delta}_{j,l}$ at
different $l$ can be considered as statistically independent
measurements. In other words, each FFC can be treated as independent
realizations of the stochastic variable $\tilde{\delta}_{j,l}$.
Thus, the FFCs, $\tilde{\delta}_{j,l}$, on scale $j$ form an ensemble with
$2^j$ realizations. The statistics with respect to the one-point
distribution of FFCs $\tilde{\delta}_{j,l}$ from one-realization
should be equal to the results of the ensemble statistics. The
goodness of this estimation is measured by the Large Number Theorem, that is,
the relative error is about $1/\sqrt{2^j}$. The one-point
distribution of FFCs from one-realization will be valid in detecting
statistical features if density field $\delta(x)$ is ergodic: the
average over an ensemble is equal to the spatial average taken over one
realization.

A homogeneous Gaussian field with continuous spectrum is certainly
ergodic (Adler 1981). More importantly, it has also been found that in
some non-Gaussian cases, such as homogeneous and isotropic turbulence
(Vanmarke, 1983), ergodicity also approximately holds. When one
considers that the density field of the universe is homogeneous and
isotropic, the one-point distributions of FFCs should be effective in
measuring the non-Gaussianity of cosmic distributions $\delta(x)$.

\bigskip

\noindent{\it 2.4 Spectrum of non-Gaussianity}

\bigskip

To take advantage of the $\tilde{\delta}_{j,l}$'s ability to detect
non-Gaussian behavior at different scales, $j$, we define the
spectrum of skewness as
\begin{equation}
S_j \equiv \frac{1}{N_{r} 2^j \sigma^3_j} \sum_{n=1}^{N_r} \sum_{l=0}^{2^j-1}
[(\tilde{\delta}_{j,l} - \overline{\tilde{\delta}_{j,l}})^3]_n,
\end{equation}
and the spectrum of kurtosis as
\begin{equation}
K_j \equiv \frac{1}{N_{r} 2^j \sigma^4_j} \sum_{n=1}^{N_r} \sum_{l=0}^{2^j-1}
[(\tilde{\delta}_{j,l} - \overline{\tilde{\delta}_{j,l}})^4]_n - 3,
\end{equation}
where the variance $\sigma^2$ is given by
\begin{equation}
\sigma^2_j= \frac{1}{2^j} \sum_{l=0}^{2^j-1}
(\tilde{\delta}_{j,l} - \overline{\tilde{\delta}_{j,l}})^2.
\end{equation}
Eq.(12) is equal to $(L/2^j) P^{var}_j$ and $P^{var}_j$ used in PF2
as the spectrum of the perturbation.

Note that eqs.(10) and (11) differ slightly from usual definition of the
skewness or kurtosis by the sum over $n$. This is because at small $j$ an
individual sample covering the range $L$ will yield only a small number of
$\tilde{\delta}_{j,l}$. This makes the calculation of $K_j$ meaningless.
For instance, for $j = 2$, there are only two FFCs,
$\tilde{\delta}_{2,0} \, \mbox{and} \, \tilde{\delta}_{2,1}$. In this
case, $\overline{\tilde{\delta}_{j,l}} = (\tilde{\delta}_{2,0} +
\tilde{\delta}_{2,1})/2 $, which using the usual definition would
yield $K_j = -2 $ regardless of the sample or wavelet.
In order to overcome this difficulty,
we compile subsets consisting of $N_r$ samples. The number of
$\tilde{\delta}_{j,l}$ will then be $N_r$ times larger than one sample
making the statistics at small $j$ viable.  As with the usual
definitions of skewness and kurtosis, $S_j$ and $K_j$ should vanish for a
Gaussian distribution.

\bigskip

\noindent{\bf 3. Demonstration of non-Gaussian detection}

\bigskip

\noindent{\it 3.1 Normal perturbations}

\bigskip

To ensure that eqs. (10) and (11) yielded the
expected results for the Gaussian case,
a 1-D density distribution $\rho(x)$ was generated from Gaussian
 perturbations with the following spectrum
\begin{equation}
P(k)=\frac{k}{1+10^5k^4}
\end{equation}
where $k=2\pi n/L$, $L$ being the range of the density field. The
spectrum (13) has a peak at $\log k \sim -1.37$, or
a typical scale at $1 /k= 23.4$ (length) units.

Samples of distributions over $L = 512$ bins were produced which gave a
bin size of 2$\pi$ units. The reconstruction
of the spectrum (13) is shown in Figure 1a. The peak and the amplitude
of the power spectrum are perfectly detected by the wavelet SSD (PF2).
The results of $S_j$ and $K_j$ are shown in Figures 1b and 1c. The
error bars are given by $\sqrt{15/N}$ and $\sqrt{96/N}$, respectively,
where $N$ is the total number of wavelet coefficients, i.e., the number of
realizations times $2^j$.
(Press et al. 1992). As expected for  normal perturbations,
both $S_j$ and $K_j$ are zero.

\bigskip

\noindent{\it 3.2 Distribution of clumps}

\bigskip

Let us consider non-Gaussian density fields consisting of clumps randomly
distribute in a white noise background. Clump distributions are often used
to test methods of detecting non-Gaussianity in large scale structure
study (Perivolaropoulos 1994, Fan \& Bardeen 1995). In fact, these kinds of
studies have shown that one
cannot detect the non-Gaussianity
of samples by the one-point probabilities of the individual Fourier modes,
even when the samples contain only a few independent clumps (Kaiser \&
Peacock 1991).
More importantly,
these distributions are necessary in studying whether random samples contain
non-Gaussian signatures (see \S 3.3).

To begin, first note that a clump or valley
with density perturbation $\Delta \rho_c$ on length scale
$d$ at position $l$ can be described as
\begin{equation}
\rho^{\pm}(x) =
   \left\{ \begin{array}{ll}
      \pm \Delta \rho_c & {\rm if} \ \ lL/2^{J_c} \leq x < (l+1)L/2^{J_c} \\
         0 & {\rm otherwise}
         \end{array}
   \right.
\end{equation}
where $J_c=\log_2 (L/d)$, and the positive sign is for a clump, the
negative sign for a valley.
If a density field $\rho(x)$ consist of $N$ randomly distributed
clumps and valleys of
scale $d$, so that the number density is $N/2^{J_c} d$
on average, the field can be realized by a random
variable of density perturbation $\delta \rho$ with a probability
distribution $P(\delta \rho)$ defined as
\begin{equation}
P[x \leq X]= \left\{ \begin{array}{ll}
                     0           & {\rm if} \ \ X < -\Delta \rho_c \\
                     N/2^{J_c+1} & {\rm if} \ \ -\Delta \rho_c < X < 0 \\
                     1-N/2^{J_c} & {\rm if} \ \ 0 < X < \Delta \rho_c \\
                     1           & {\rm if} \ \ X > \Delta \rho_c
                    \end{array}
           \right.
\end{equation}
The distribution function $\delta \rho$ of clumps
and valleys, $f_c(\delta \rho)$ can then be written approximately as
\begin{equation}
f_c(x)= \frac{dP}{dx}=
(1-\frac{N}{2^{J_c}})\delta(x) +
\frac{N}{2^{J_c+1}}\delta(x -\Delta \rho_c)  +
\frac{N}{2^{J_c+1}}\delta(x +\Delta \rho_c).
\end{equation}
The $\delta(..)$ on the right hand side of eq.(16) denote
$\delta$-functions.
The characteristic function of the random variable $\delta \rho$
of clumps and valleys is
\begin{equation}
\phi_c(u)=\int_{-\infty}^{\infty} f(x)e^{i \delta u} dx
=\frac {2^{J_c}-N}{2^{J_c}} + \frac {N}{2^{J_c}} \cos(\Delta_c u)
\end{equation}
where $\Delta_c=\Delta \rho_c/\bar \rho$.
It is very well known that the overall measures of skewness and
kurtosis of the distribution (15) can be calculated from the
characteristic function (17). The results are
\begin{equation}
S = - \frac{1}{i\sigma^3}\left [\frac{d^3\phi_c(u)}{du^3} \right ]_{u=0}=0
\end{equation}
and
\begin{equation}
K = \frac{1}{\sigma^4}\left [\frac{d^4\phi_c(u)}{du^4} \right ]_{u=0} - 3
  = \frac{2^{J_c}}{N}-3.
\end{equation}
where
\begin{equation}
\sigma^2= - \frac{d^2\phi_c(u)}{du^2}|_{u=0}
= \frac {N (\Delta_c)^2}{2^{J_c}}
\end{equation}
is the variance of the distribution.

Consider density fields consisting of clumps or valleys randomly
distributed in a background. In this case, the characteristic
function is $\phi(u) = \phi_c(u) \phi_b(u)$, where $\phi_b(u)$ is the
characteristic function of the background distribution. For a
randomly uniform Gaussian
background with variance $\sigma_b^2$, the overall variance is
\begin{equation}
\sigma^2= \frac {N (\Delta_c)^2}{2^{J_c}} + \sigma_b^2,
\end{equation}
and the overall kurtosis is
\begin{equation}
K = \left ( \frac{2^{J_c}}{N} -3 \right )
\left (1+\frac{2^{J_c}}{N (s/n)^2} \right)^{-2},
\end{equation}
where $s/n=\Delta_c /\sigma_b$ is the signal-to-noise ratio. Eq.(22)
shows that this distribution becomes Gaussian when $s/n$ is small.

Samples of clumps and valleys randomly distributed were produced with a
Gaussian noise background.
Figure 2a shows a typical distribution. Here, $d = 1$ bin, there are
16 clumps and valleys on average, and the signal-to-noise ratio  $s/n = 5$.
The size of the  distribution is $ L = 512$ bins.
The spectra, $S_j$ and $K_j$, are
shown in Figures 2b and 2c. Since the distributions are non-Gaussian, a
Gaussian variance will no longer be applicable to estimate the statistical
errors. Instead, the error bars in Figure 2 are calculated from the 95\%
confidence level from the ensemble of the samples (for each
parameter set at least 100 realizations are generated).
The kurtosis calculated from the one point function of the Fourier
modes has also been
plotted in Figure 2. This figure clearly shows that non-Gaussian behavior
can be detected by the FFCs, but not by Fourier methods.

Figure 3 shows the kurtosis spectrum for a distribution consisting
of 16  randomly distributed clumps and valleys with scale $d =4$ bins,
over a length $L=512$ bins, and $s/n = 5$. $K_j$ is significantly
different from zero on scale $j=6$, which corresponds exactly to
the scale of the clumps. i.e. $512/2^{6+1}=4$. (The $+1$ in the index is
due to the way scaling is counted for the FFC's.  See eq. (8) and figure 1
of PF1.)  This demonstrates that the scale of the
clumps can be detected by the peak in the spectrum of kurtosis.

The effectiveness of detecting multiple scales of the clumps and valleys
has also been tested.  Figure 4 shows the results of generating
samples consisting of 16, 32, and 48 clumps and valleys with a $s/n = 2.0$,
and the scales of the clumps, $d$, are randomly distributed
from 1 to 5 bins. Once again 100 realizations are generated. The kurtosis
spectra are plotted in Figure 4. Also shown is the
standard one value description of kurtosis, plotted for clarity of
presentation at $j = 9$.  This kurtosis is directly calculated from the
distribution $\delta(x)$ by
\begin{equation}
K \equiv \frac{1}{N 2^j \sigma^4} \sum_{i=1}^{N}
[(\delta(x_i) - \overline{\delta(x_i)})^4] - 3.
\end{equation}

Several features stand out in the Figure 4.  First,
the standard singled valued kurtosis is generally lower
than that given by the
wavelet kurtosis, especially, when the number of clumps is large. In
this case
the standard kurtosis totally misses the non-Gaussianity of the distribution.
 Second,
the one value kurtosis contains a large uncertainty in detecting
deviations from Gaussian behavior. This is because the distribution
$\delta(x)$ is equal  to about the
MFCs on finest scale. As mentioned in \S 2.2, MFC's distributions
will be Gaussian if the clumps are independent on various scales.
On the other hand, the kurtosis spectrum detected the non-Gaussian signal
at $j = 6$ even when the number of clumps is as large as 48, and
$s/n = 2.0$. Since the average bin width is about 3, this
corresponds to about 1/3 of the 512 bins being occupied by clumps.
The FFC's are extremely sensitive to deviations from
Gaussian behavior.

Figure 5 directly plots the one-point distributions of the FFCs for
realizations with sixteen 4d clumps and valleys over a length 512d, and
with $s/n$=5. In each panel, the Gaussian distribution (dotted line) with
the same variance and normalization as the FFC distribution is shown.
The figure shows that the largest deviation of the distribution of FFCs
from Gaussian distribution occurs on the scales of the clumps ($j = 6$).

\bigskip

\noindent{\it 3.3 Non-Gaussianity of random samples}

\bigskip

In large scale structure study, the usual way of generating random
distributions
covering a 1-dimensional range ($x_1,x_2$) is
\begin{equation}
x_i = x_1 + (x_2 - x_1)\cdot RAN
\end{equation}
where $x_i$ is the position of i-th object, and $RAN$ is random
number in (0, 1). Because the number of objects is an integer,
the random samples given by eq.(24) easily lead to non-Gaussian distributions.

In numerical calculations, the distribution $\delta(x)$ is often binned into
a histogram with a given bin size. If the bin size is
less than the mean distance of neighbor objects, the value of the
binned $\delta(x)$ will typically be 0 or 1. The sample is then a d=1 clump
distribution with a one point distribution given by eq.(15), and
{\em not} a Gaussian distribution. Only in the case when the mean
 number of objects
contained in one bin is large does the one-point distribution approach the
Gaussian case. To illustrate this point, Figure 6 plots the spectrum of
kurtosis for a sample generated by eq.(24), in which the number of objects
is 122 distributed in 64 bins. The figure shows that the
non-Gaussianity of the random sample is significant when the mean number
of objects/bins is equal to 2.

Similarly, the binning of real data will lead to non-Gaussianity that is
obviously not in the original distribution. Generally, in order to maximally
pick up information from a real data set, the bin size is taken to be the
resolution of the coordinate $x$, i.e. the lowest possible
size of the binning. However, this data reduction will also lead the
lowest mean ratio of objects per bin and so consequently lead to the highest
non-Gaussianity.

\bigskip

\noindent{\bf 4. Non-Gaussian detection in the Ly$\alpha$ forests}

\bigskip

\noindent{\it 4.1 Simulation samples of Ly$\alpha$ forests}

\bigskip

The non-Gaussian behavior of samples given by simulations of the
Ly$\alpha$ forests (Bi, Ge \& Fang 1995, hereafter BGF) were examined. These
samples have also been
used for the demonstration of cluster identification and spectrum analysis
by a wavelet SSD (PF1, PF2). The details of the simulations
are given by BGF. The basic steps of the simulations are as follows:
1) generate dark matter distributions by Gaussian
perturbations with a linear power spectrum of the standard cold dark
matter model (SCDM), the cold plus hot dark matter model (CHDM),
and the low-density flat cold dark matter model (LCDM); 2) generate the
baryonic matter distribution by assuming that baryonic matter traces the
dark matter distribution on scales larger than the Jeans length of the
baryonic gas, but is smooth over structures on scales less than the Jeans
length; 3) remove collapsed regions from the density field because
Ly$\alpha$ clouds are probably not virialized; 4) simulate Ly$\alpha$
absorption spectrum as the absorption of neutral hydrogen in the baryonic
gas, and include the effects of the observational instrumental
point-spread-function, and along with Poisson and background noises;
5) determine the Ly$\alpha$ absorption line and its width from the simulated
spectrum by the usual way of Ly$\alpha$ line identification.

Within a reasonable range of  the UV background radiation
at high redshift, and the threshold of the onset of
gravitational collapse of the baryonic matter, the LCDM model is consistent
with observational features including 1) the number density of Ly$\alpha$
lines and its dependencies on redshift and equivalent width; 2) the
distribution of equivalent widths and its redshift dependence; 3)
two-point correlation functions; and 4) the Gunn-Peterson effect.
Especially important is the fact that the simulated data show no
power in the two-point
correlation function and that their 1-dimensional spectra is flat on
scales less than 100 h$^{-1}$Mpc (BGF, PF2).

However, no power in the two-point correlation and/or a flat spectrum
does not mean that the sample is white noise. Instead, this may only
indicate that the power spectrum and two-point correlation function are not
suitable for describing the statistical features of the system being
considered.
Indeed, using a multi-resolution SSD analysis, the distributions of
Ly$\alpha$ forest samples with no power in the two-point correlation function
have been found to be significantly different from uniformly random
distributions on various scales (PF1). The non-Gaussianity spectrum analysis
will support this result.

As in PF1 and PF2, a one dimensional distribution $n(x)$ of the Ly$\alpha$
lines was formed by writing each sample into a histogram with
bins on comoving scale of 2.5 h$^{-1}$ Mpc. This is about the distance at
which the effect of line
blending occurs. We then generated 100 uniformly
random samples for each simulation sample via eq. (24). Since the lines are
redshift
dependent the total number of lines and the number of lines within a given
red-shift interval (say, $\bigtriangleup z = 0.4$) of the random samples
were chosen to match the parent distribution. We calculated the FFCs
and MFCs for both the BGF sample and the random data. The original
distribution $n(x)$ and its various scales of multiresoluted results
can perfectly be reconstructed by the MFCs (PF1).

Figure 7 gives the one-point distribution of FFCs for a SCDM sample
(W$>0.36 \AA$).
For each scale $j$, the corresponding Gaussian distribution is plotted such
that it has the same normalization and variance as the one-point
distribution. This figure clearly shows that all the distributions on
$j>6$ (or scales less than about 40 h$^{-1}$ Mpc) are significantly
non-Gaussian. It is interesting to point out that the BGF simulation is
based on a linear spectrum, and the perturbation field is Gaussian. The
detected non-Gaussian behavior must come from the selection of high peaks
in the Gaussian background fields (step 5 of the simulation procedure).

Figures 8 and 9 are the spectra of skewness and kurtosis of the SCDM
samples (W$>0.36 \AA$). In order to test whether the detected
 non-Gaussianity is from
the binning, Figures 8 and 9 also show the spectra of skewness and
kurtosis for two random samples, I and II. Random II is produced from
the Gaussian one-point distributions, which have the same normalization and
variance in each $j$ as the SCDM sample (Figure 7). As expected, the kurtosis
and skewness of Random I are equal to zero. Random I
is generated using eq. (24). The redshift-dependence of the number
density of the Ly$\alpha$ lines is accounted for by generating the random
data such that in each redshift range the number of lines in the
random sample is the same as the SCDM sample. Hence, the spectra of the
Random II samples is a measure of the possible non-Gaussianities due to the
binning.

Figure 9 shows that the amplitude of the kurtosis spectrum for the
SCDM sample is systematically larger than the corresponding Random I sample.
Recall that the error bars in Figure 9 do not represent the 1 $\sigma$
Gaussian errors, but the 95\% confidence level from the ensemble of the
samples. The difference between the spectra of the SCDM and
Random I samples is significant.

The skewness in Figure 8 is small, consistent with zero, but slightly
positive. Even the Random I data has small, but positive skewness. A
possible reason for the positive skewness is the redshift-dependence
of number of Ly$\alpha$ clouds. The FFC
$\tilde{\delta}_{j-1,l}$ is mainly determined by the
difference of (positive) densities of $\{j, 2k\}$ and $\{j, 2k+1\}$ (PF1).
Namely, for a clump in redshift space, the density change on the lower
redshift or lower $k$ side contributes negative FFCs, while the higher
redshift side give positive FFCs. If the number of Ly$\alpha$ clumps
decreases with increasing redshift, the change in clustering amplitudes
(FFCs) on the higher redshift side (positive FFCs) should be less than
the lower side (negative FFCs), i.e. the number of positive
FFCs will be less than negative FFCs. In Figure 7, one can see the
asymmetry (non-zero skewness) in the one-point distributions.

Figure 10 and 11 give the skewness and kurtosis spectra for samples
of all three models (W$>0.16\AA$). For all models, the skewness is about the
 same however the
kurtosis is different for different models.  For the CHDM data, the $K_j$
amplitudes are larger than that of the SCDM and LCDM data on all scales $j$.
This is because there are far fewer high peaks in the CHDM than
in SCDM and LCDM.  The kurtosis is therefore a useful measure in
distinguishing between the various models.

\bigskip

\noindent{\it 4.2 Real data}

\bigskip

As in the first two papers (PF1 and PF2), two data sets of the Ly$\alpha$
forests are examined. The first was compiled by Lu, Wolfe and Turnshek
(1991, hereafter LWT).
It contains $\sim$ 950 lines from the spectra of 43 QSOs that
exhibit neither broad absorption line nor metal line systems. The second
is from Bechtold (1994, hereafter JB), which contains a total $\sim$
2800 lines from 78 QSO's spectra, in which 34 high redshift QSOs were
observed at moderate resolution. In our statistics, the effect of
proximity to $z_{em}$ has been considered. All lines with redshift
$z_{em} \geq z \geq z_{em} - 0.15$ were deleted from our samples. We
assumed $q_{0} = 1/2$, so the distance of the samples range from a
comoving distance from about $D_{min}$=2,300 $h^{-1}$Mpc to
$D_{max}=$3,300 $h^{-1}$Mpc.

A problem in using real data to do statistics is the complex geometry of
QSO's Ly$\alpha$ forest. Different forest cover different
spatial ranges, and no one of the forests distributes on the entire range
of $(D_{min}, D_{max})$. This is a difficulty in detecting the power
spectrum by some of the usual methods. At the very least, a complicated
weighting
scheme is needed. However, for the wavelet SSD, this problem can easily
be solved. For instance, suppose a forest sample lies in a range
$(D_1, D_2)$. The range can be extended to $(D_{min},D_{max})$ by
adding zeros to the data in ranges $(D_{min}, D_1)$ and $(D_2,D_{max})$.
Since the father functions are compactly supported,
the FFCs in the range $(D_1, D_2)$ will not be affected by the
addition of zero in the  ranges of $(D_{min},D_1)$ and $(D_2, D_{max})$.
Any statistic can then be computed by simply dropping all FFCs,
$\tilde\psi_{j,l}$, with coordinates $l$ in the added zero ranges.
The only uncertainty due to the boundary at $(D_{min},D_{max})$
are the two FFCs at boundary, $\tilde\psi_{j,l_1}$ and
$\tilde\psi_{j,l_2}$, where $l_1$ and $l_2$ correspond to the positions
in the regions of
$(D_{min}, D_1)$ and $(D_2,D_{max})$. In PF2, we  numerically
demonstrated that FFCs can correctly detect the ``local" spectrum, i.e.
spectrum in the range $(D_1, D_2)$, regardless of any zero padding.
In other words, the
FFC spectrum reconstruction is insensitive to the selection of boundary
conditions. Using this technique, all samples were extended in
comoving space, to cover 1024 bins with each bin of comoving
size $\sim 2.5$ h$^{-1}$ Mpc. Thus, all QSO samples were treated
uniformly.

Another problem in computing the spectrum of skewness and kurtosis of
real data is the compilation of the subsets of the sample, needed
for eqs.(10) and (11). Unlike the simulated samples, where as much data as
needed can be generated, the available real data are limited. There are
only $N_T=$ 43 samples (forests) for LWT and 78 for JB. In order to
effectively use this data, $M \leq N_T$ files from among the complete $N_T$
samples are chosen to form a subset.  Various combinations of the subsets $M$
are then combined to form an ensemble.  To investigate the effect of
different
combinations, the subsets $M$ are formed by varying the total number
of files chosen from
the parent distribution, $N_T$, as well as changing the order in which the
individual files are selected.
It is found
that the skewness and kurtosis calculated from these $M$-file ensembles
are very stable until $M$ contains few as 7 or 8 files, i.e.
until only approximately
5\% of the total lines remain in the subset. The 95 \% confidence
intervals are then estimated from the ensembles.

Figure 12 gives the FFC one-point distributions for sample of
LWT $> 0.36 \AA$ and JB $>0.32 \AA$. As in Figure 7, these distributions
show highly non-Gaussian behavior, and are also asymmetric, with fewer
positive FFC's and more negative ones.

The skewness and kurtosis spectra of both the LWT and JB data are
plotted in Figure 13 and 14, respectively. Even though the two data sets are
independent they give the same amplitudes for the kurtosis
on all scales $j>6$ (or larger than 40 h$^{-1}$Mpc). Therefore, these
amplitudes are not be statistical flukes, but come from the real
clustering of Ly$\alpha$ clouds. Figure 14 also show the kurtosis of
samples SCDM and LCDM. In terms of second and lower order statistics,
such as number density and two-point correlation functions, LCDM gives
best fitting of observed data. However, LCDM's kurtosis are found to
be systematically less than real data.

To check for non-Gaussianity due to binning, Figures 15 and 16 give
the skewness and kurtosis spectra of the JB (W$>0.16 \AA$) sample and
its Random I data. The Random I samples have the same number
density as the Ly$\alpha$ lines in each redshift ranges $\Delta z =0.4$ as
the real data. The difference of kurtosis between the JB and its Random I
sample (Figure 16) is more significant than the difference between SCDM
and its Random I data (Figure 9). This indicates that real data is more
non-Gaussian than the linear simulation sample.

As in Figure 8, all the skewness of LWT, JB and JB's Random I are
consistent with zero, but slightly positive. This is probably because
the positive evolution of the number of Ly$\alpha$ clusters with
redshift.

\bigskip

\noindent{\bf 5 Discussion and conclusions}

\bigskip

We have demonstrated that the one-point distribution of FFCs of
the discrete wavelet is a good tool for detecting non-Gaussian behavior in
cosmic density fields. The locality of the wavelet farther functions
in both configuration and Fourier spaces allows for a way around the
central limit theorem:  a superposition of randomly distributed
non-Gaussian clumps will be Gaussian.

It is generally believed that the ergodic hypothesis is reasonable if spatial
correlations are decreasing sufficiently rapidly with increasing separation.
In this case, volumes separated with distances larger than the correlation
length can be considered as statistically independent regions. Therefore,
in terms of short-range correlated components, such volumes can be treated
as independent realizations. Many theoretical models indeed predict that
the perturbations in the universe are short-range correlated, or, at least,
that the universe contains short-range correlated components.
The wavelet FFCs effectively measure these statistically independent
regions. Thus, in the case where the ``fair sample hypothesis"
(Peebles 1980) holds, the FFC one-point distribution taken from
one-realization is
a fair estimate of the PDF of density fields.

The spectra of skewness and kurtosis provide a systematic and uniform
measure of the  non-Gaussianity of various samples. This method is
sensitive to
samples containing many clumps embedded in a Gaussian background, while
Fourier methods fail to do so (Perivolaropoulos 1994).  As
opposed to the PDF given by CIC, which is statistically incomplete,
the FFCs give a complete description of the scale-dependence of the skewness
and kurtosis.

The spectra of skewness and kurtosis have been detected for QSO's Ly$\alpha$
forests in both observational and linearly simulated samples.
In previous studies these samples have shown no power in the
 two-point correlation function and
a flat power spectrum. The results of the non-Gaussian detection are
non-trivial: all distributions are found to be non-Gaussian on at
least scales
of less than 40 h$^{-1}$Mpc, and the
non-Gaussianities are not completely due to the effects of binning.
It is clear that high order statistics, such as the spectrum
of the kurtosis, can indeed provide information which is missed by the
2nd order statistics.
The amplitude and shape of the kurtosis spectrum are found
to be the same for the two independent data sets. Thus, the features
shown in the kurtosis spectrum should come from the formation and evolution
of Ly$\alpha$ clouds. We will study the dynamical implication of these
features in future work.

The kurtosis spectrum for the BGF simulation samples of Ly$\alpha$ forests
was also calculated. The kurtosis spectra are different for different
dark matter models describing the formation of Ly$\alpha$ clouds. Among the
BGF
samples, the best fitting to the observed number density and its evolution
of Ly$\alpha$ clouds is given by the LCDM data. However, the kurtosis
spectrum of the LCDM sample is significantly lower than real data. This
result is consistent with that given by cluster identification. In PF1, it
was
found that the ratio between the numbers of larger and lower scale
clusters for the real data is greater than that of the LCDM data.
Obviously, the larger the cluster number ratio, the larger the deviation
from a Gaussian distribution. Hence, the kurtosis and skewness spectrum
opens a new window for looking at the statistical features of large scale
structures. It is an important addition to the existing methods of
describing the clustering and correlation of the cosmic density field, and
for discriminating among models of structure formation.

\bigskip

Both authors wish to thank Professor P. Carruthers, and Drs.H.G. Bi and P.
Lipa for insightful conversations.

\bigskip

\newpage

\noindent{\bf References}

\bigskip

\ownref Adler, R.J. 1981, {\it The Geometry of Random Field}, (New York, Wiley)

\ownref Alimi, J.M., Blanchard, A. \& Schaeffer, R. 1990, ApJ,
  402, 38

\ownref Bouchet, F., Strauss, M.A., Davis, M., Fisher, K.B., Yahil, A.
   \& Huchra, J.P. 1993, ApJ, 417, 36

\ownref Bechtold, J. 1994, ApJS, 91, 1.

\ownref Bi, H.G., Ge, J. \& Fang, L.Z. 1995, ApJ, 452, 90, (BGF)

\ownref Daubechies, I., 1992, {\it Ten Lectures on Wavelets}, SIAM

\ownref Duncan, R.C., Ostriker, J.P. \& Bajtlik, S. 1989, ApJ, 345, 39

\ownref Fan Z.H. \& Bardeen, J.M. 1995, astro-ph/9505017

\ownref Fang, L.Z. 1991, A\&A, 244, 1

\ownref Farge, M.,1992, Ann. Rev. Fluid Mech., 24, 395

\ownref Feldman, H.A., Kaiser N. \& Peacock, J.A. 1994, ApJ 426, 23.

\ownref Gazta\~naga, E. \& Frieman, J. 1994, Astrophys. J. Lett. 437, L13

\ownref Gazta\~naga, E. \& Yokoyama, J. 1993, ApJ, 403, 4

\ownref Hamilton, A.J.S. 1985, ApJ, 292. L35

\ownref Ivanov, A.V. \& Leonenko, N.N. 1989, {\it Statistical Analysis of
Random Field},  Klumer Academic Pub.

\ownref Juszkiewicz, R., Bouchet, F.R. \& Colombi, S. 1993, ApJ,
   Lett, 412, L9

\ownref Kaiser, N. \& Peacock, J.A. 1991, ApJ, 379, 482

\ownref Kofman, L., Bertschinger, E., Gelb, M.J., Nusser, A. \& Dekel, A.
  1994, ApJ, 420, 44

\ownref Liu X.D., Jones B.J.T. 1990, MNRAS, 242, 678

\ownref Lu, L., Wolfe, A.M., \& Turnshek, D.A. 1991, ApJ, 367, 19 (LWT)

\ownref Meyer, Y. 1993, {\it Wavelets: Algorithms and Applications}, SIAM

\ownref Pando, J. \& Fang, L.Z. 1996, ApJ, 459, 1 (PF1)

\ownref Pando, J. \& Fang, L.Z. 1995, astro-ph/9509032 (PF2)

\ownref Peebles, P.J.E. 1980, {\it The Large Scale Structure of the Universe},
     Princeton Univ. Press

\ownref Perivolaropoulos, L. 1994, MNRAS, 267, 529

\ownref Press, W.H., Flannery, B.P., Teukolsky, S.A. \& Vetterling, W.T.,
     1992, {\it Numerical Recipes}, Cambridge

\ownref Suginohara, T. \& Suto, Y. 1991, ApJ, 371, 470

\ownref Vanmarke, E.H. 1983, {\it Random Field}, MIT

\ownref Yamada, M. \& Ohkitani, K. 1991, Prog. Theor. Phys., 86, 799

\newpage

\noindent{\bf Figure captions}

\begin{description}

\item[{\bf Figure 1}] Spectra of skewness and kurtosis for Gaussian
perturbations
  with perturbation spectrum $P(k)=k/(1+10^5k^4)$, which has a peak at
   $\log k \sim -1.37$, where $k=2\pi n/L$, $L$ being the range of the sample.
   The samples are produced over $L = 1024$ bins, and the bin size is 2$\pi$
   units.
   a.) the reconstructed  spectrum $P(k)$, b.) $ S_j$ and c.) $K_j$.
 The error bars are given by $\sqrt{15/N}$ for the skewness, and
  $\sqrt{96/N}$ for the kurtosis, where $N$ is the number of
  wavelet coefficients.

\item[{\bf Figure 2}] Spectra of skewness and kurtosis for clump distribution,
 which consist of sixteen $d$ clumps and valleys randomly distributed
 in a Gaussian noise background. The length of the distribution is $L=512 d$,
and
 signal-to-noise ratio $s/n$ is 5. a) A typical realization of
 the distribution; b) $S_j$; c) $K_j$; and d) the kurtosis calculated from
 one point function of FFT.

\item[{\bf Figure 3}] Spectra of skewness and kurtosis for clump distribution,
 consisting of 16 randomly distributed clumps and valleys with scale $d =4$
 bins,
 and over a length $L=512$ bins. $s/n$ is 5.

\item[{\bf Figure 4}] Kurtosis spectra for samples consisting of 16, 32, and
  48 clumps. The sizes of the clumps, $d$, are randomly distributed in range
  of 1 to 5 bins. $s/n$ is 2, and the length of the sample is
  $L=512$ bins.

\item[{\bf Figure 5}] Histogram of the one-point distribution of FFCs
for a clump distribution consisting of 16 clumps and valleys with $d=4$ bins
in range $L=512$ bins, and $s/n=5$. The vertical coordinate is relative.
At each
scale $j$, the Gauss distribution (dashed line) has the same variance
and normalization as the FFC distribution.

\item[{\bf Figure 6}] Spectrum of kurtosis of 120 objects randomly
distributed in 64 bins.

\item[{\bf Figure 7}] Histogram of the one-point distribution of FFCs
for a BGF sample in SCDM model. The width of the Ly$\alpha$ lines is
$\ge 0.36$\AA. The vertical coordinate is the number
of FFCs. At each scale $j$, the Gaussian distribution (dashed line) has the
same variance and normalization as the FFC distribution.

\item[{\bf Figure 8}] Skewness spectrum of BGF sample of SCDM model.
The solid and dashed lines are the spectra of Random I and Random II samples,
respectively.

\item[{\bf Figure 9}]  Kurtosis spectrum of BGF sample of SCDM model.
The solid and dashed lines are the spectra of Random I and Random II
samples, respectively.

\item[{\bf Figure 10}] Skewness spectrum of BGF samples in the
SCDM, LCDM and CHDM with W $\ge 0.16\AA$.

\item[{\bf Figure 11}] Kurtosis spectrum of BGF samples in the
SCDM, LCDM, and CHDM with W $\ge 0.16\AA$.

\item[{\bf Figure 12}] Histogram of one-point distribution of FFCs for
samples of LWT (W$>0.36\AA$) and JB (W$>0.32\AA$). The vertical
coordinate is relative. At each scale $j$, the Gauss distribution
(dashed line) has the same variance and normalization as the FFC
distribution.

\item[{\bf Figure 13}] Skewness spectrum for LWT and JB data.

\item[{\bf Figure 14}] Kurtosis spectrum for LWT and JB data. For
comparison, the corresponded $K_j$ of the SCDM and LCDM are also
plotted. The width of lines is $>0.16 \AA$.

\item[{\bf Figure 15}] Skewness spectra for JB data (W$>0.16\AA$),
and its Random II sample.

\item[{\bf Figure 16}] Kurtosis spectra for JB data (W$>0.16\AA$),
and its Random II sample.

\end{description}

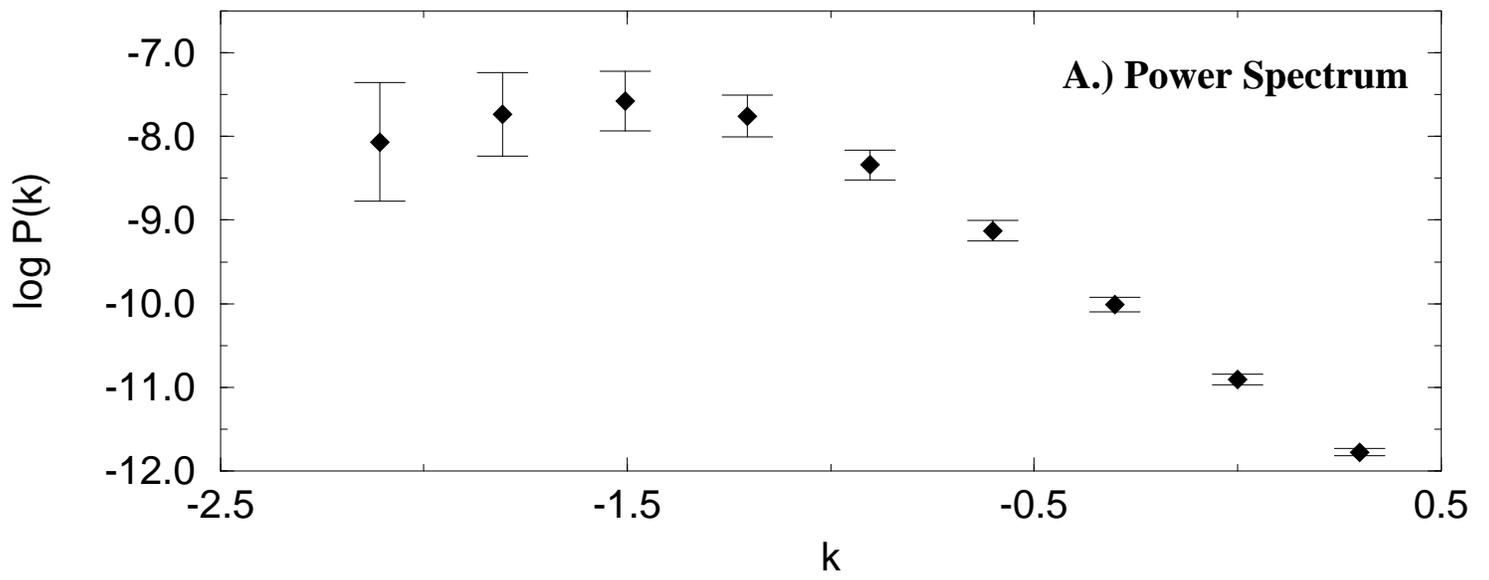
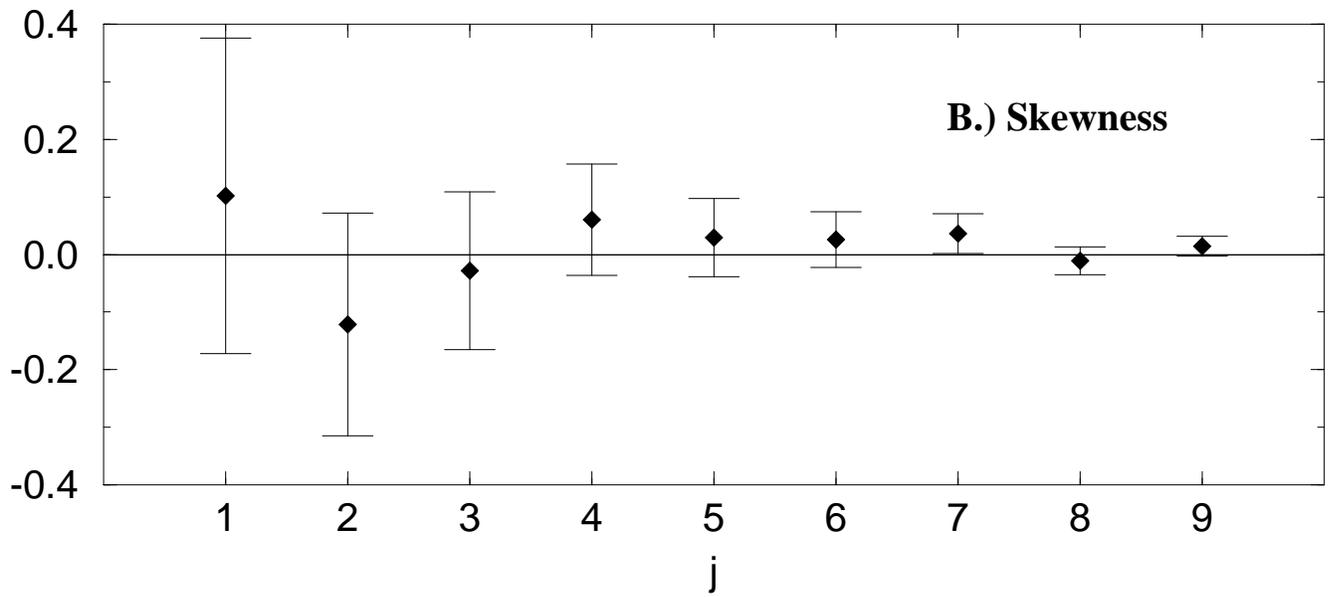
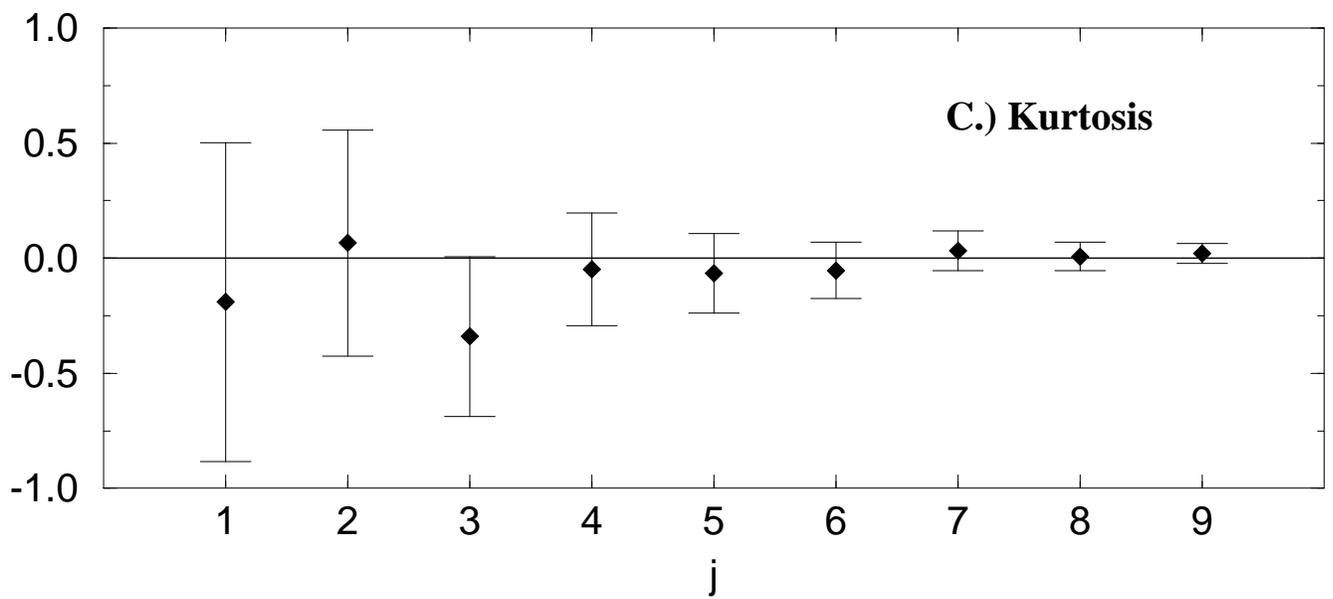
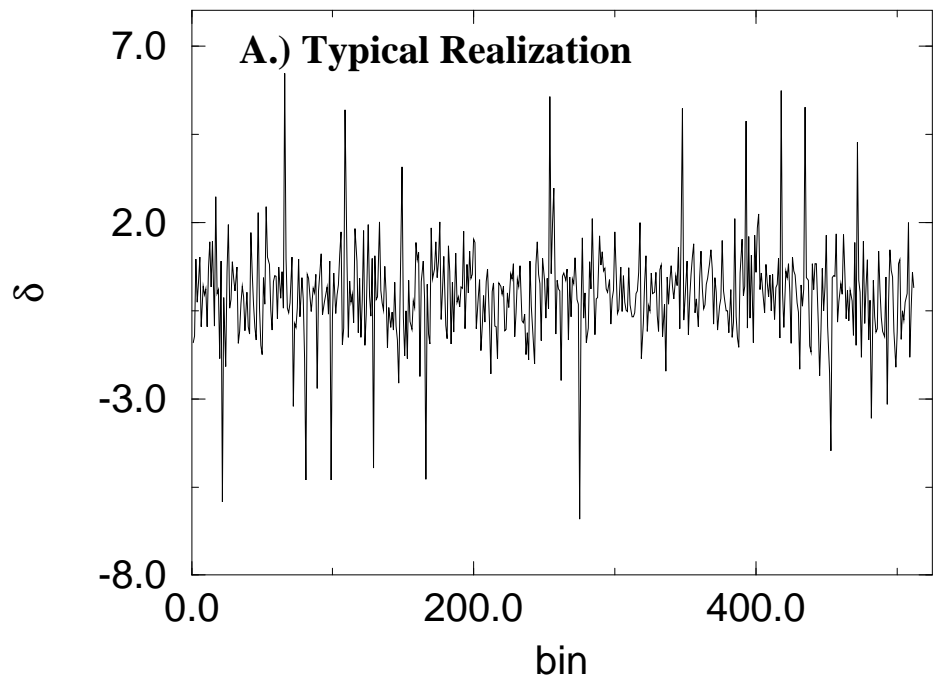
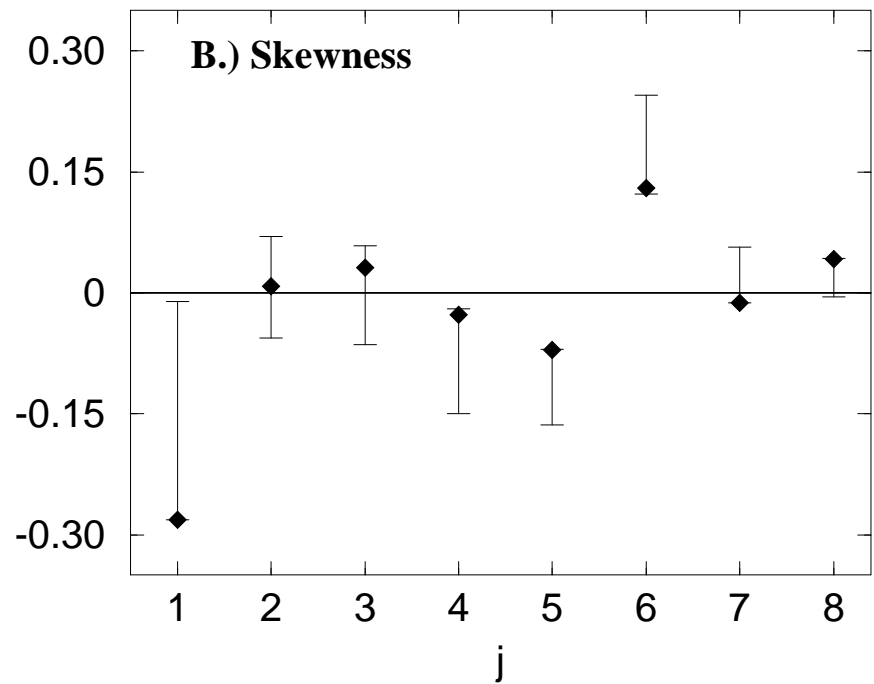
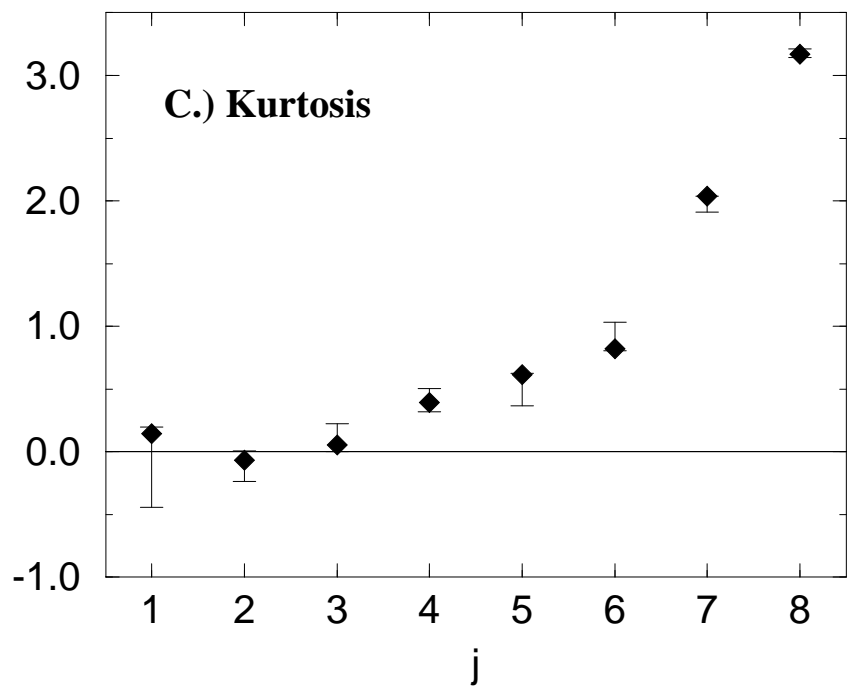
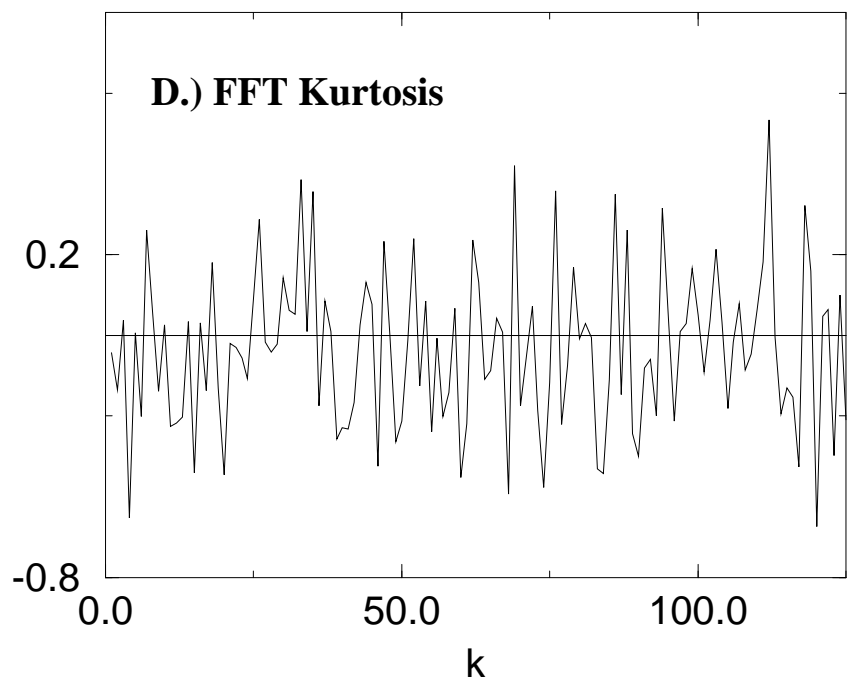
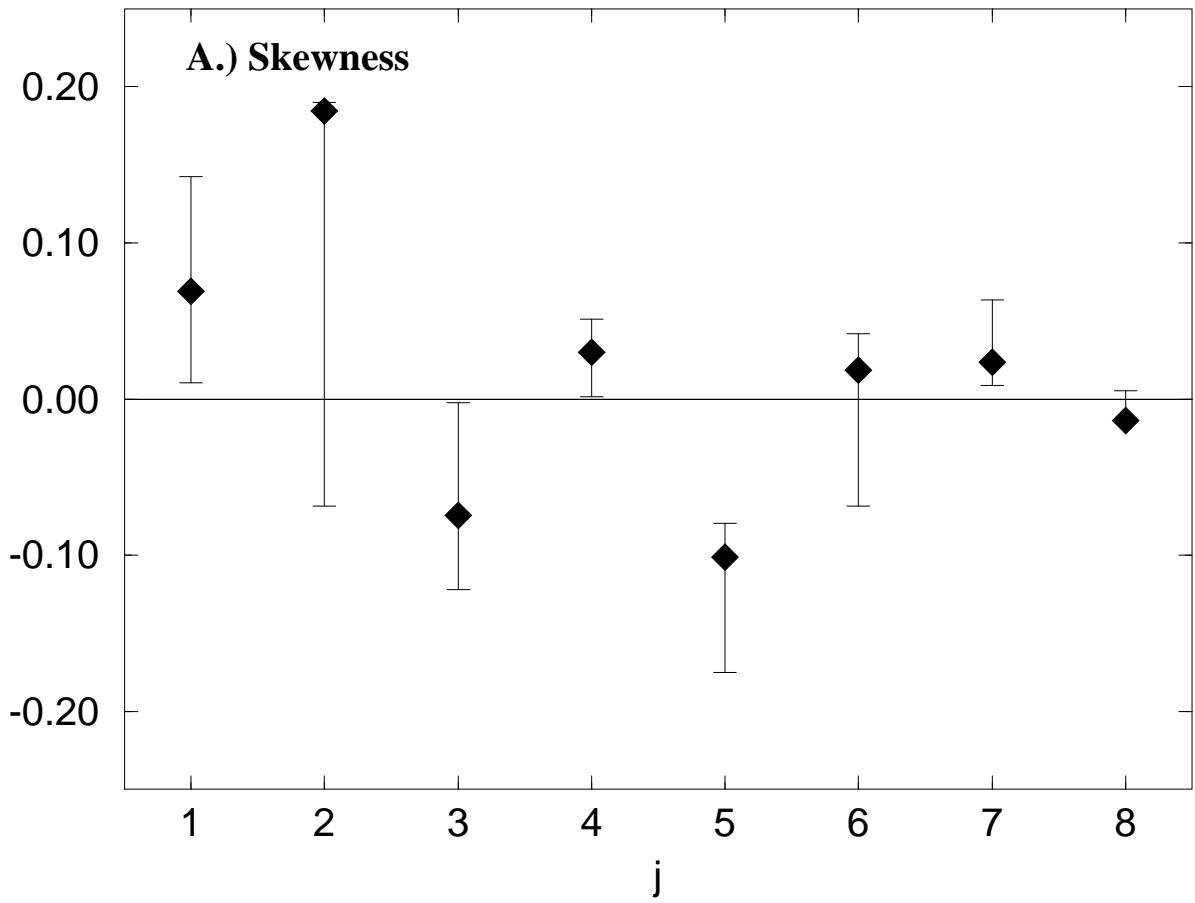
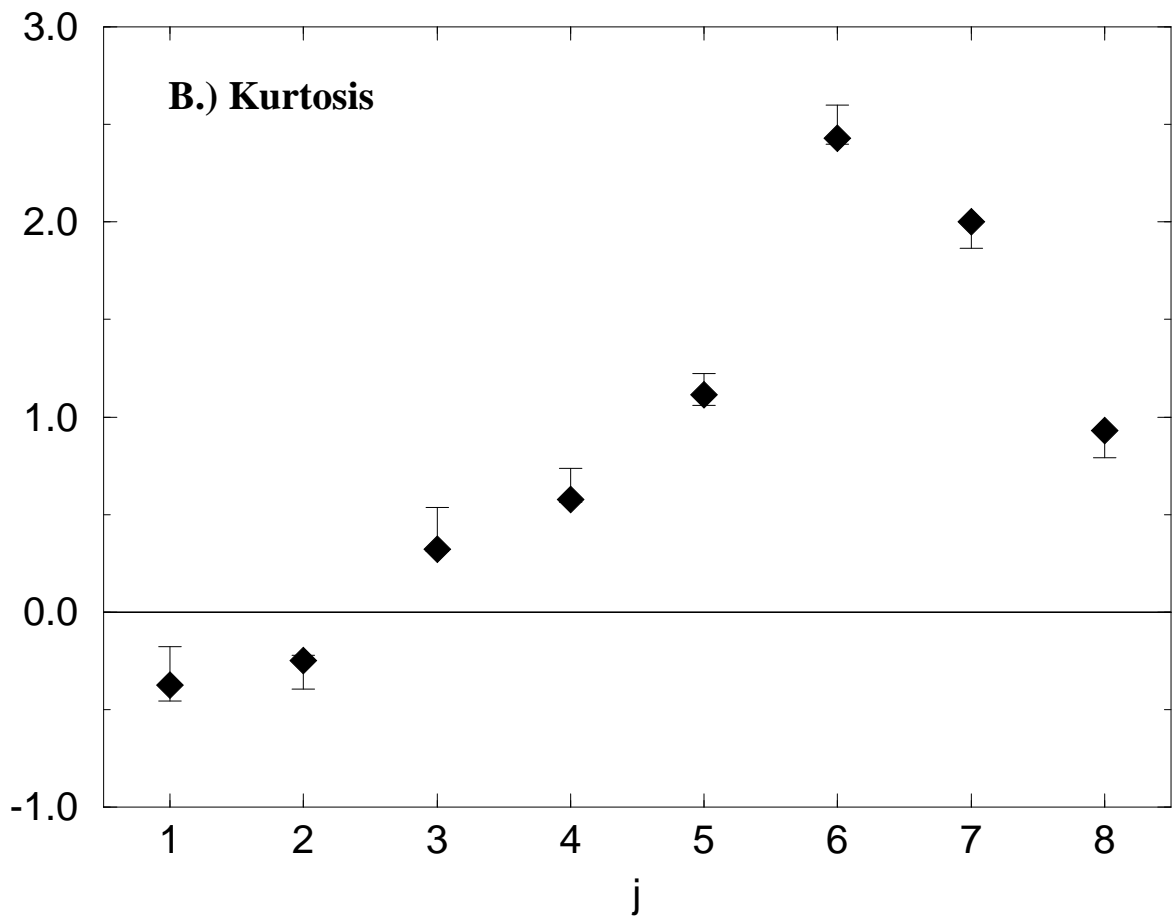
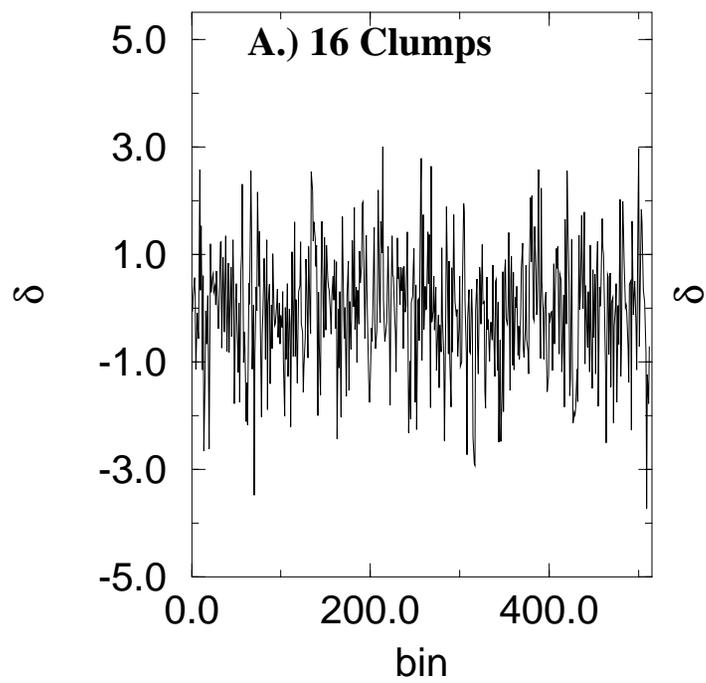
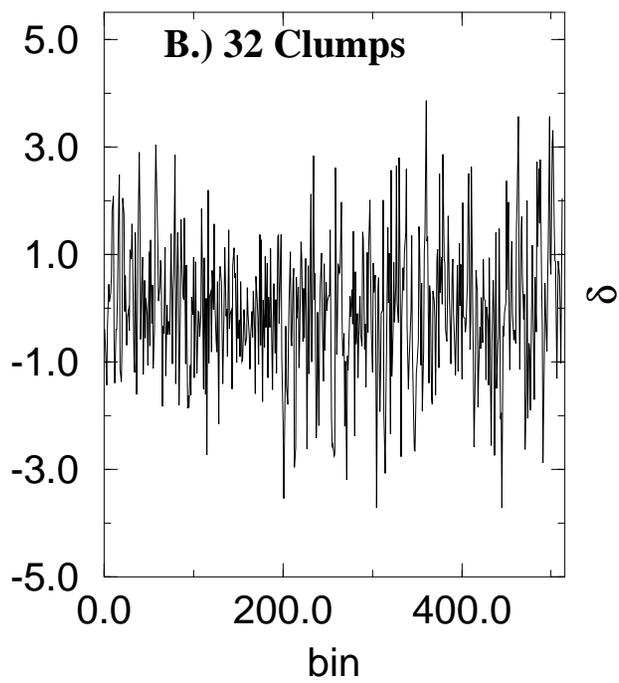
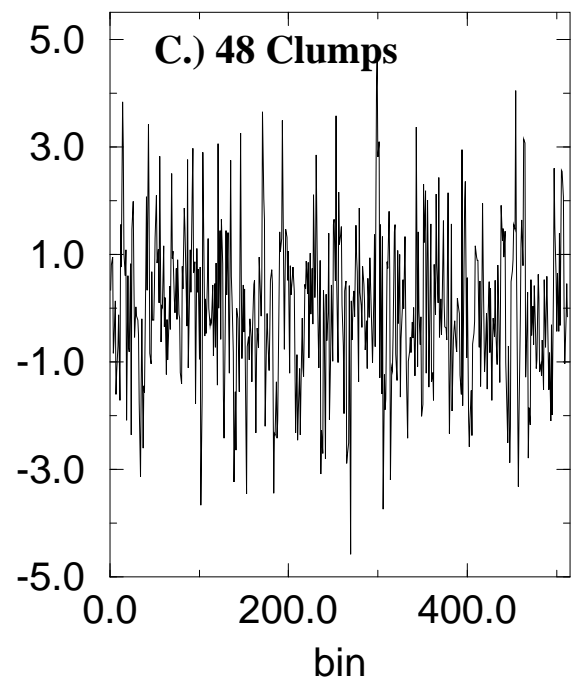
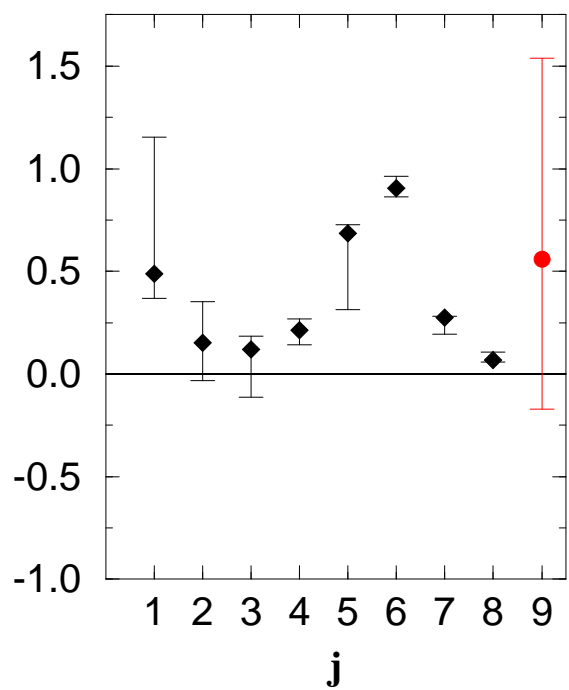
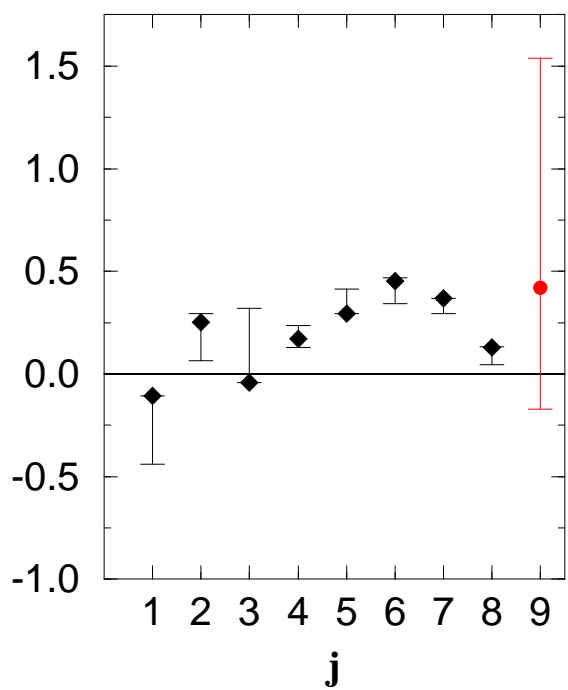
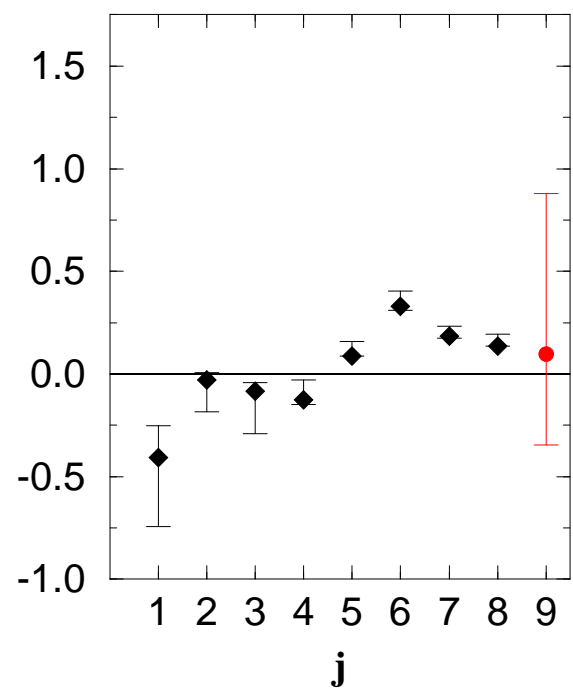
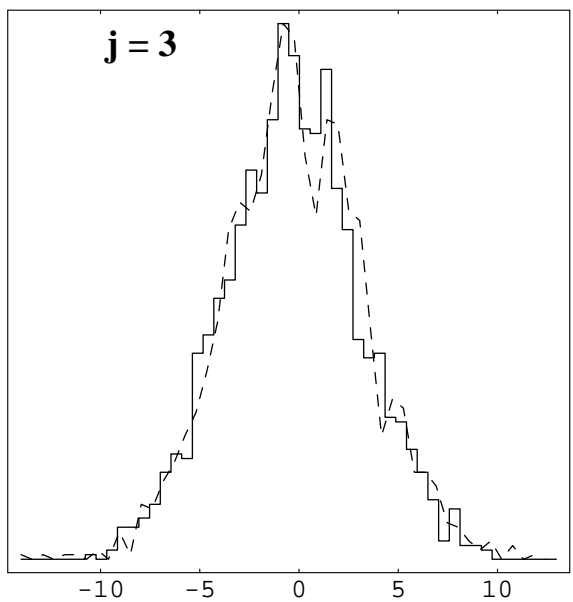
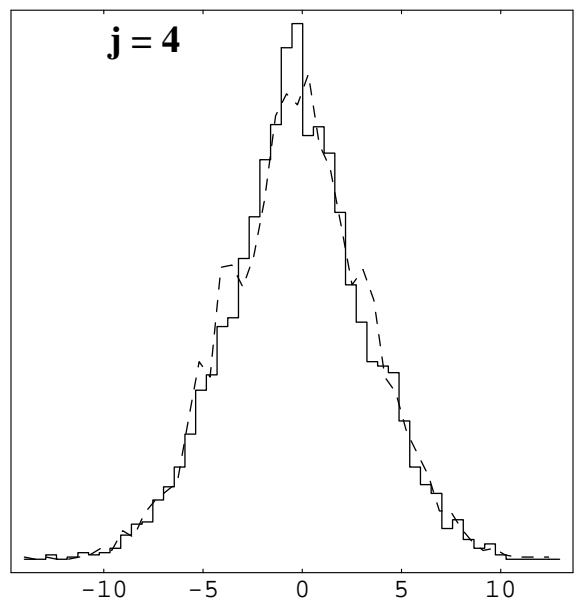
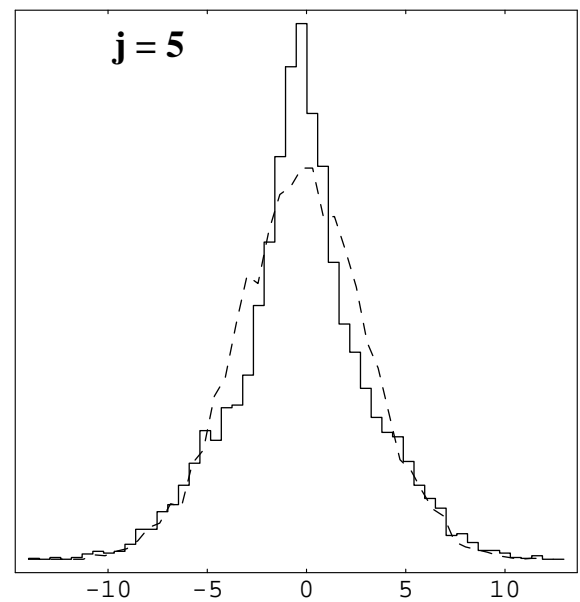
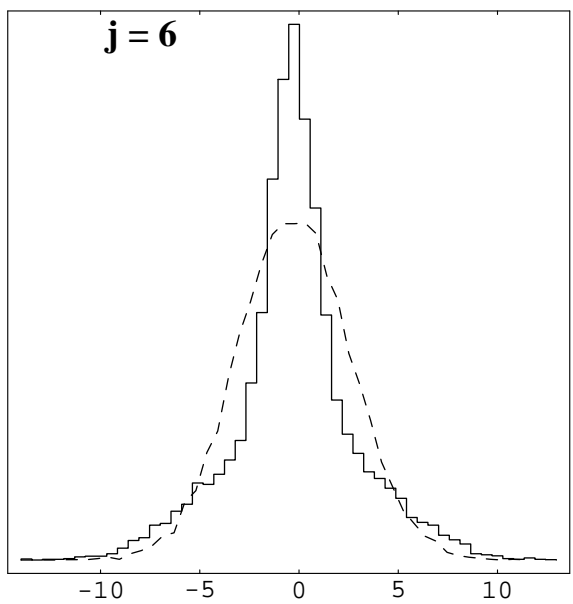
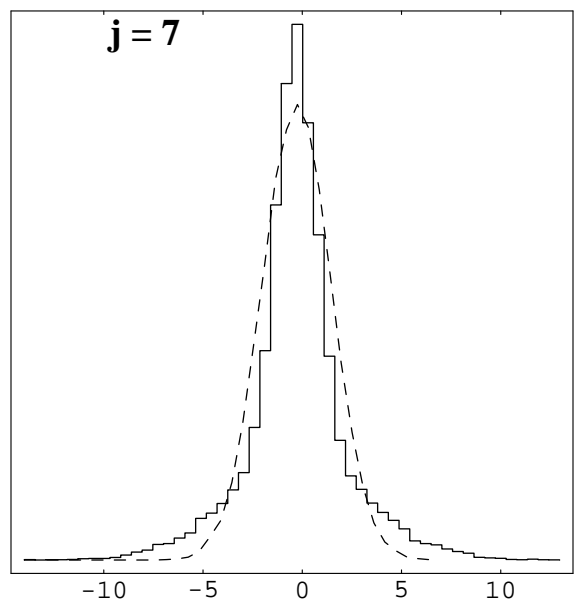
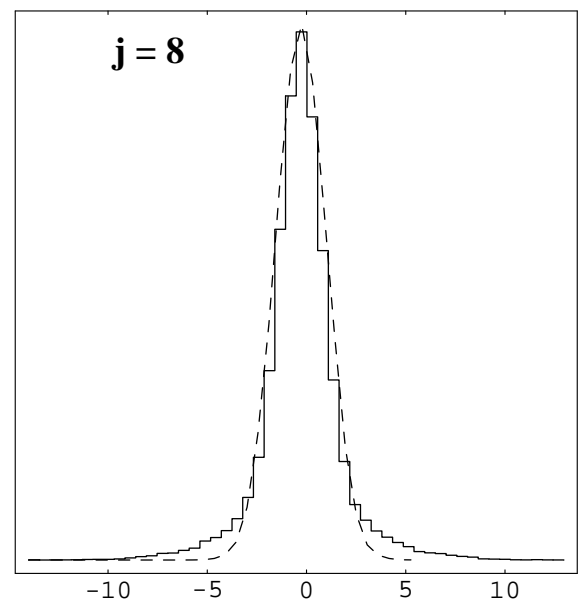
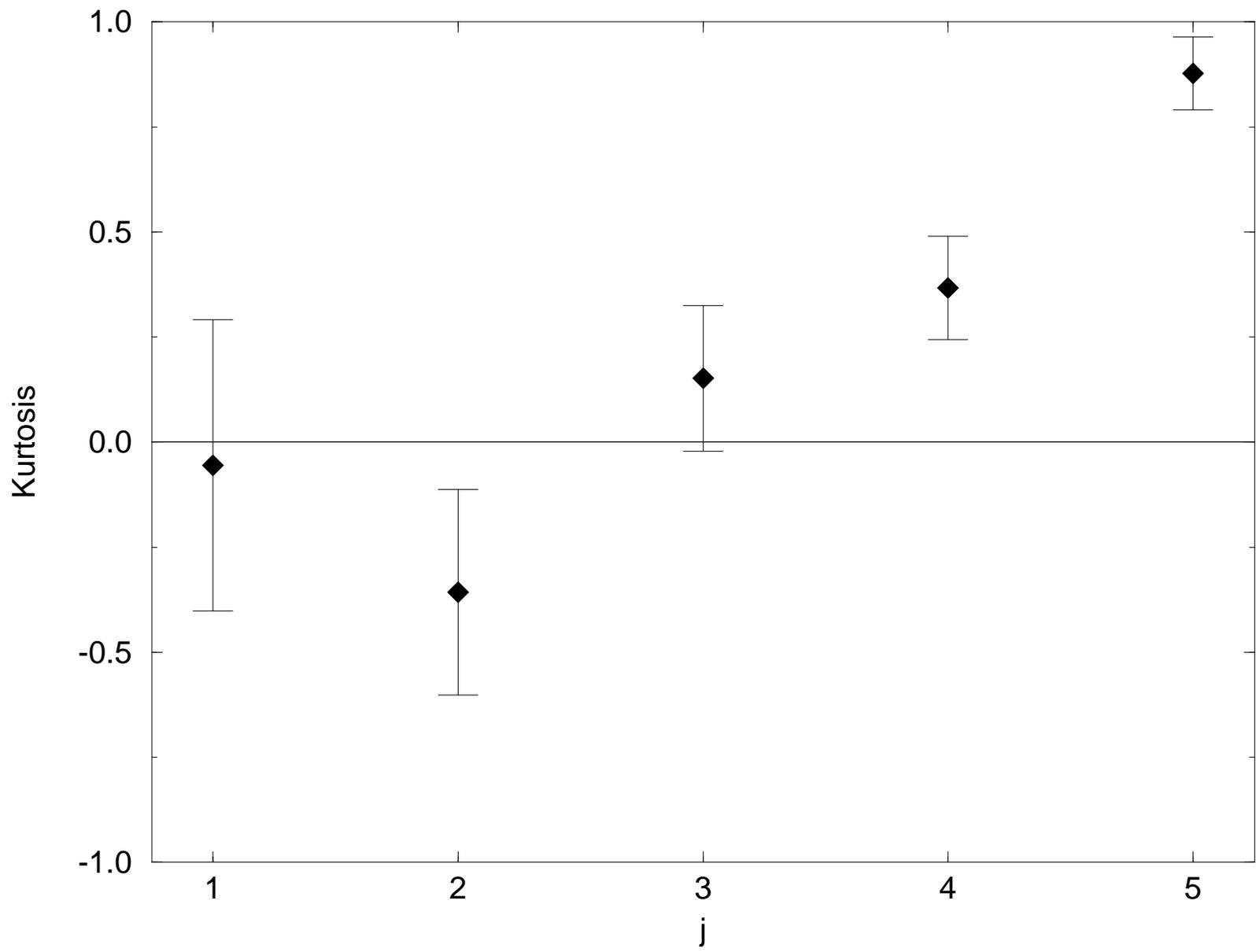
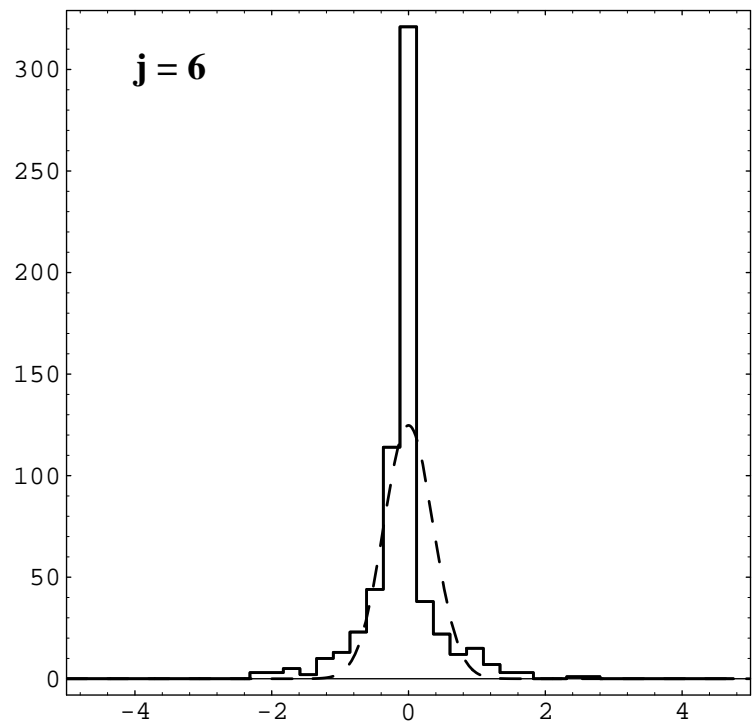
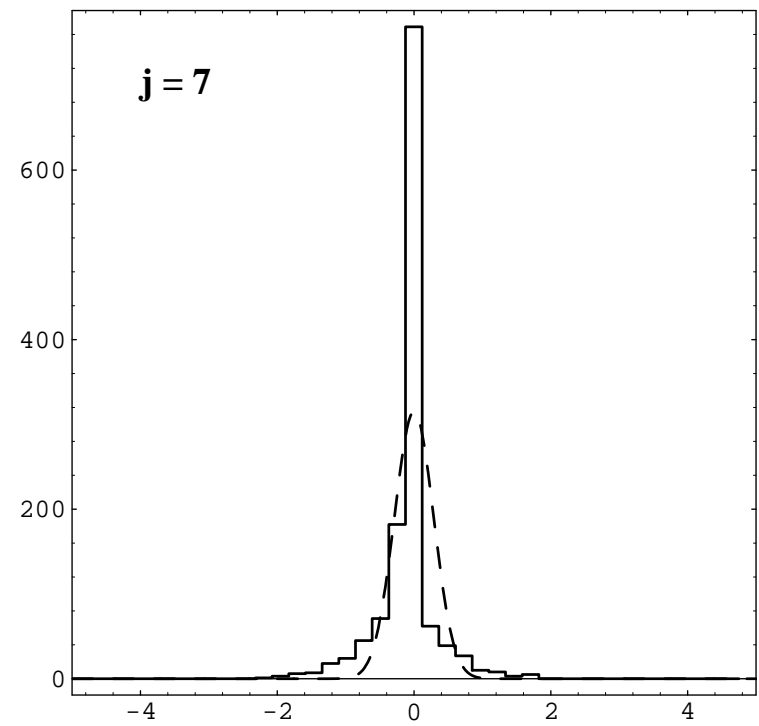
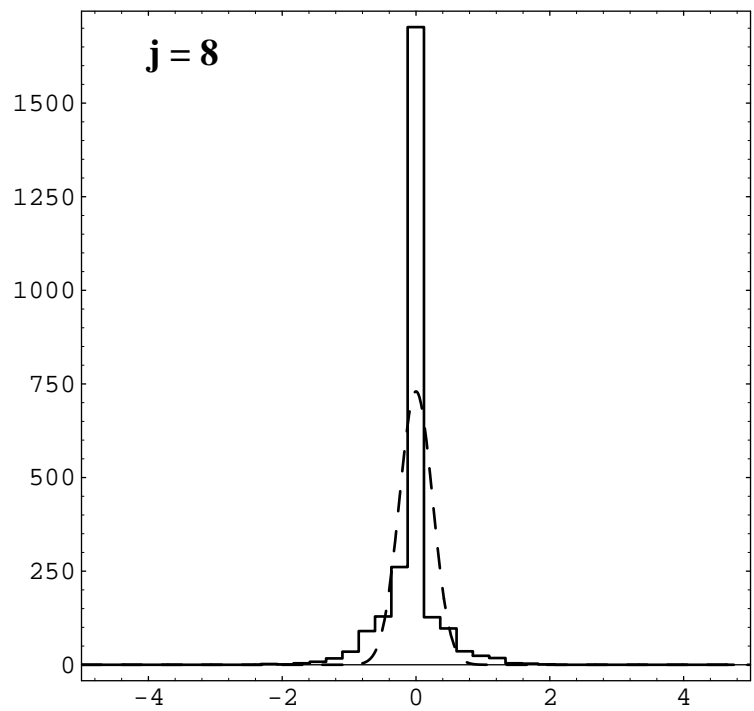
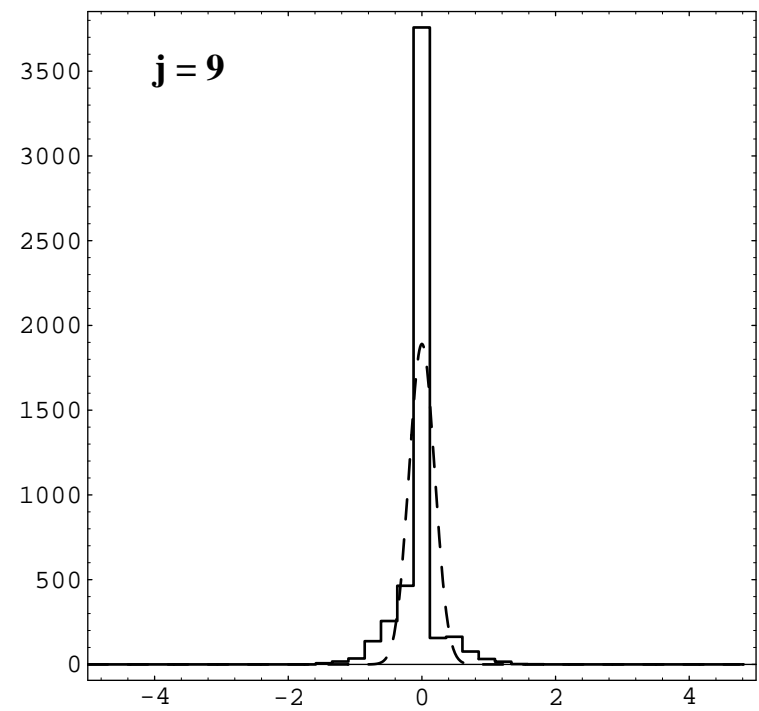
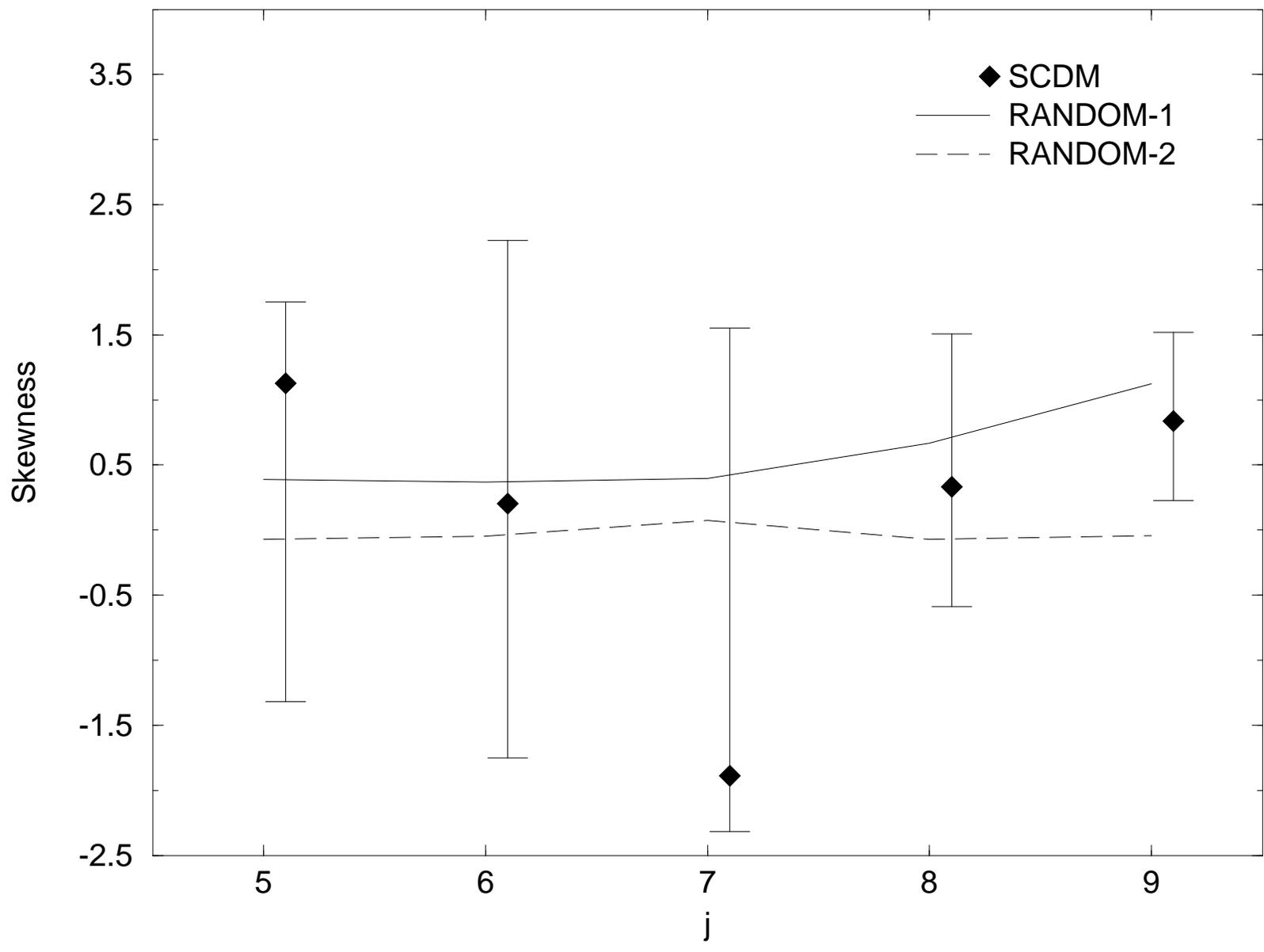
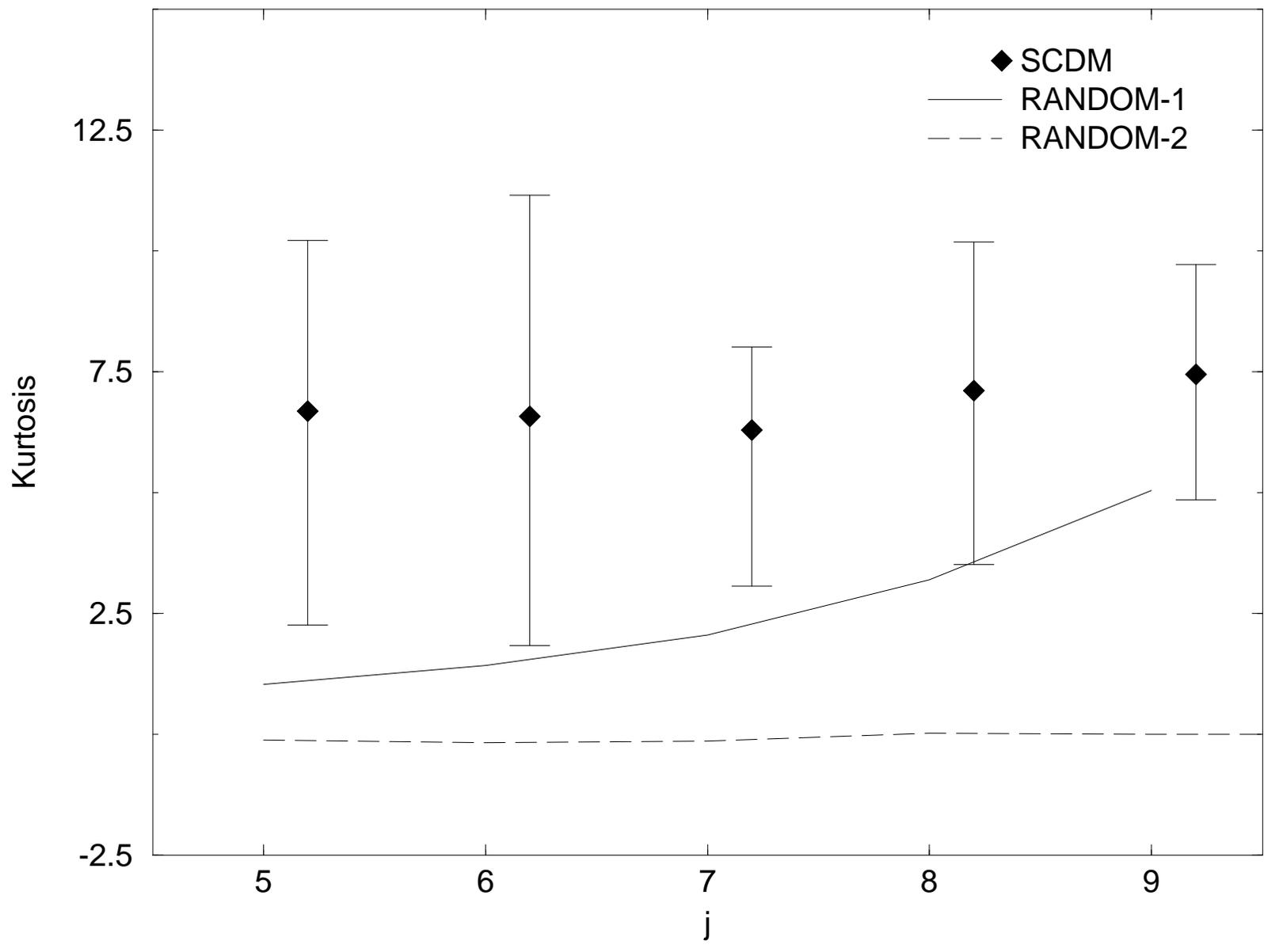
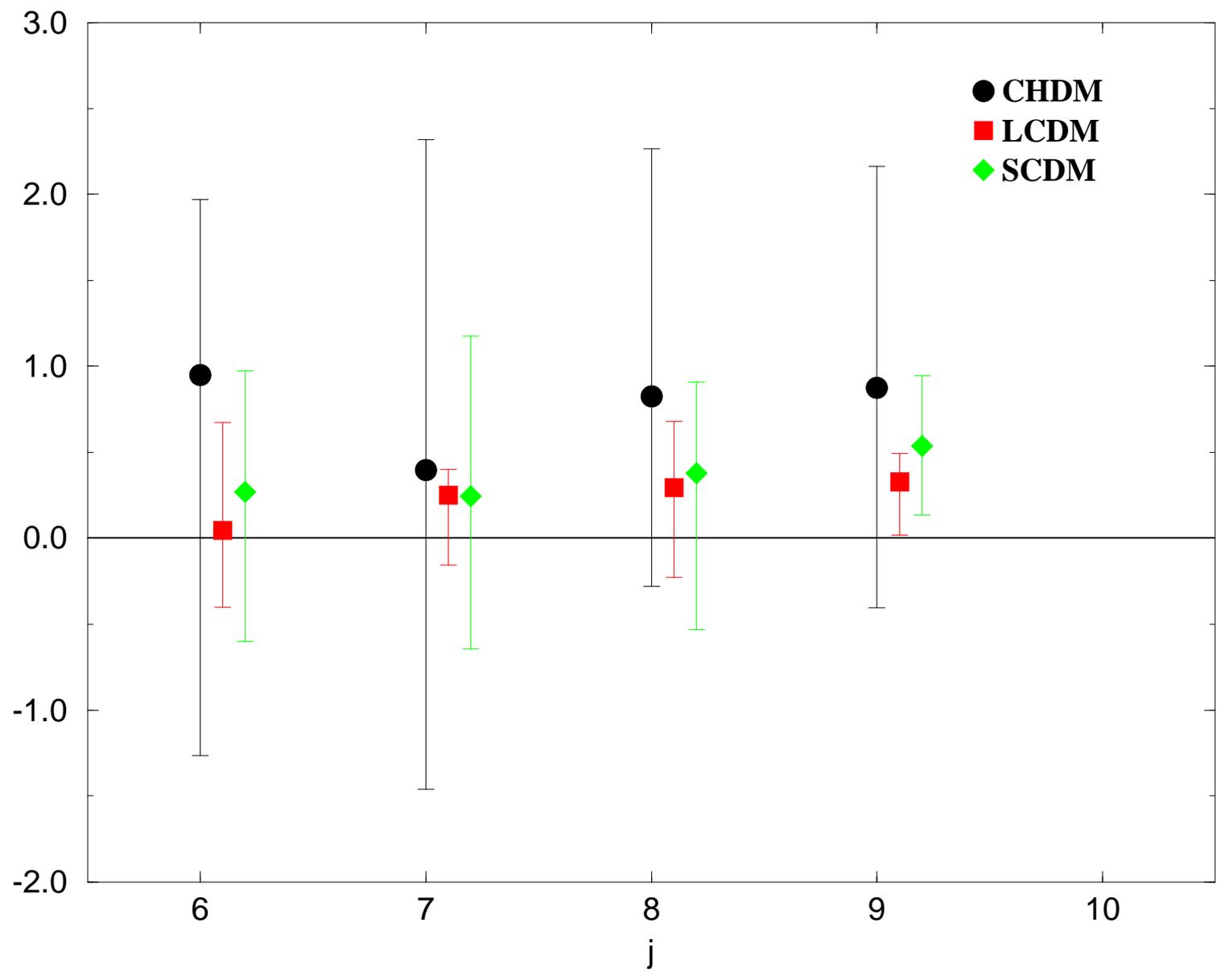
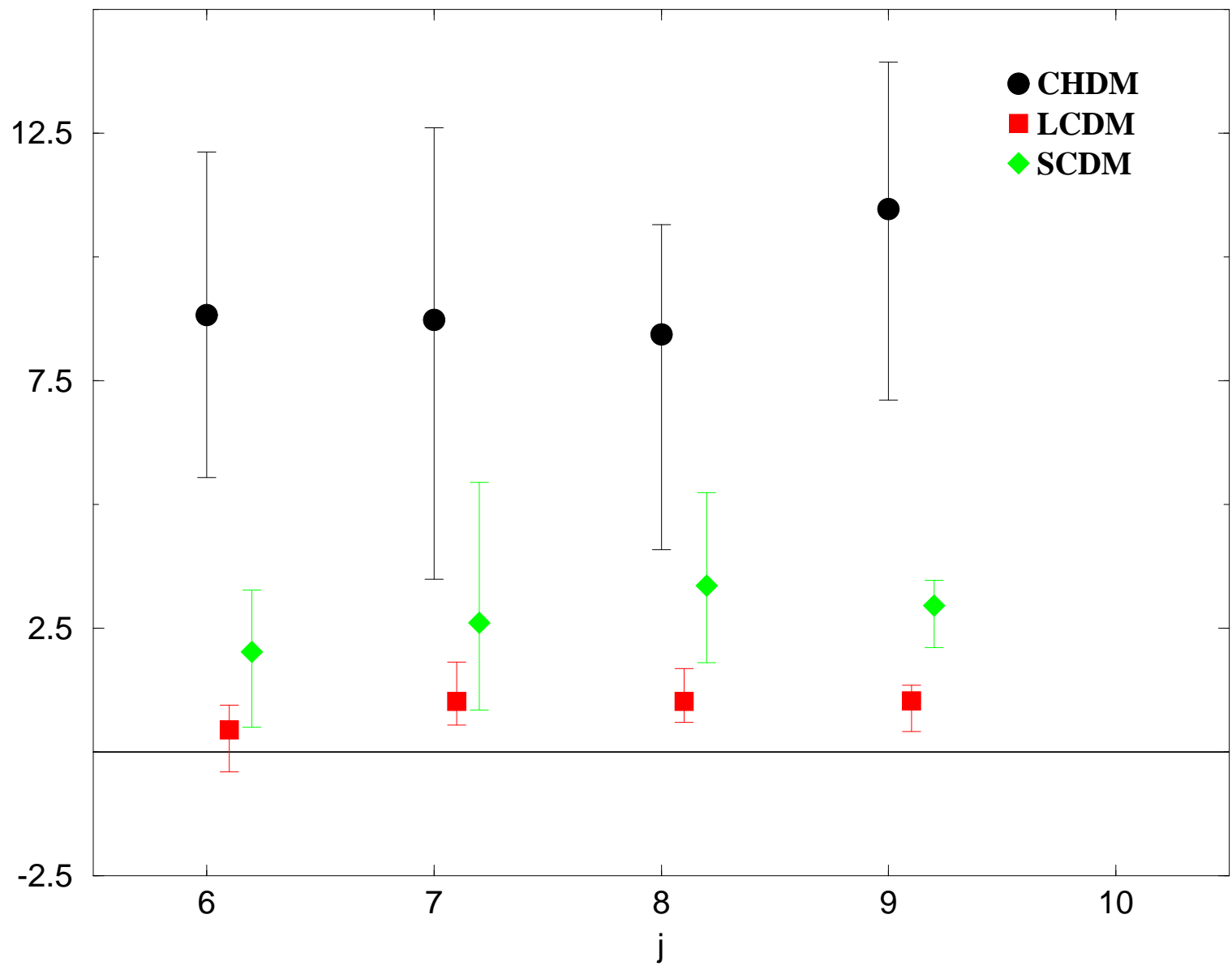
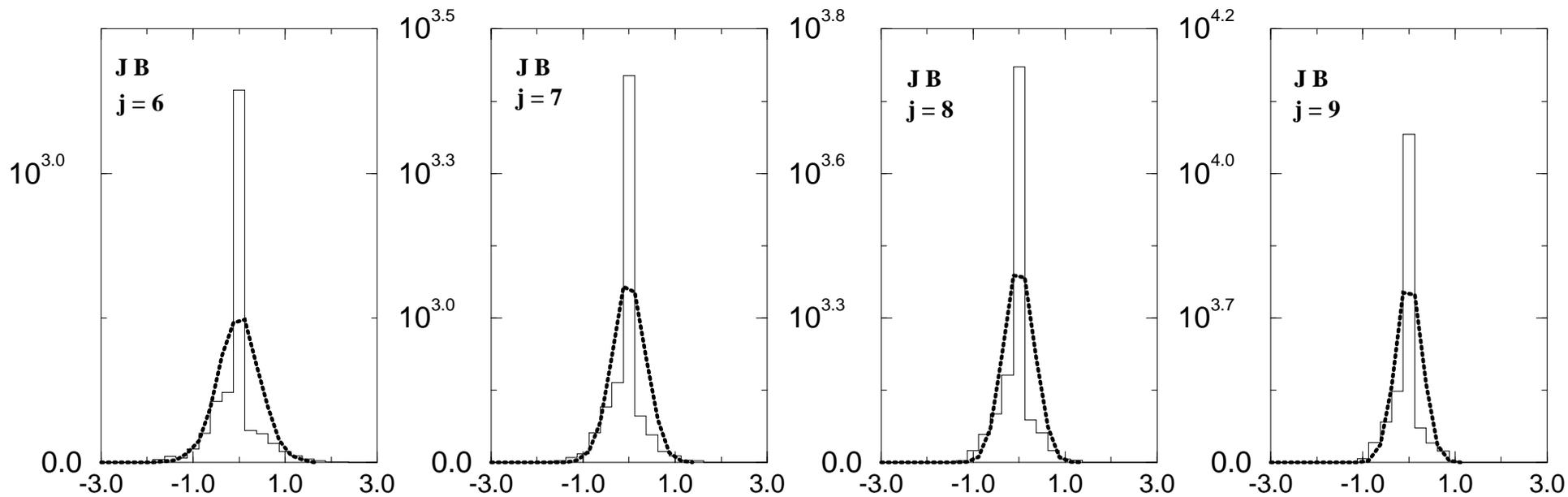
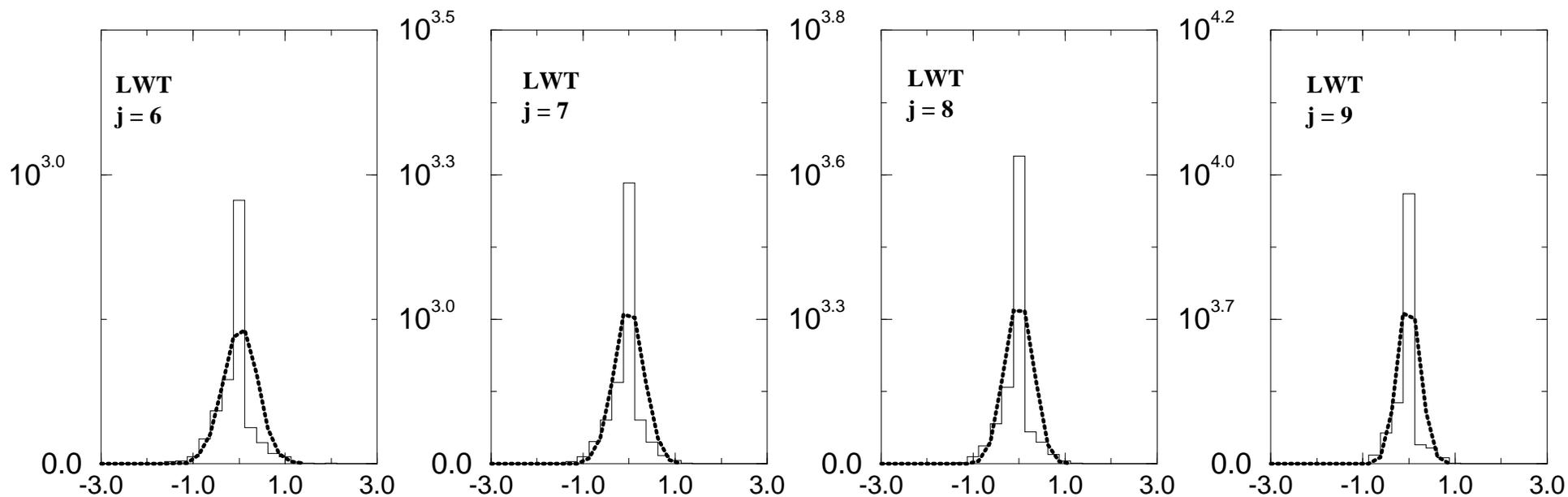
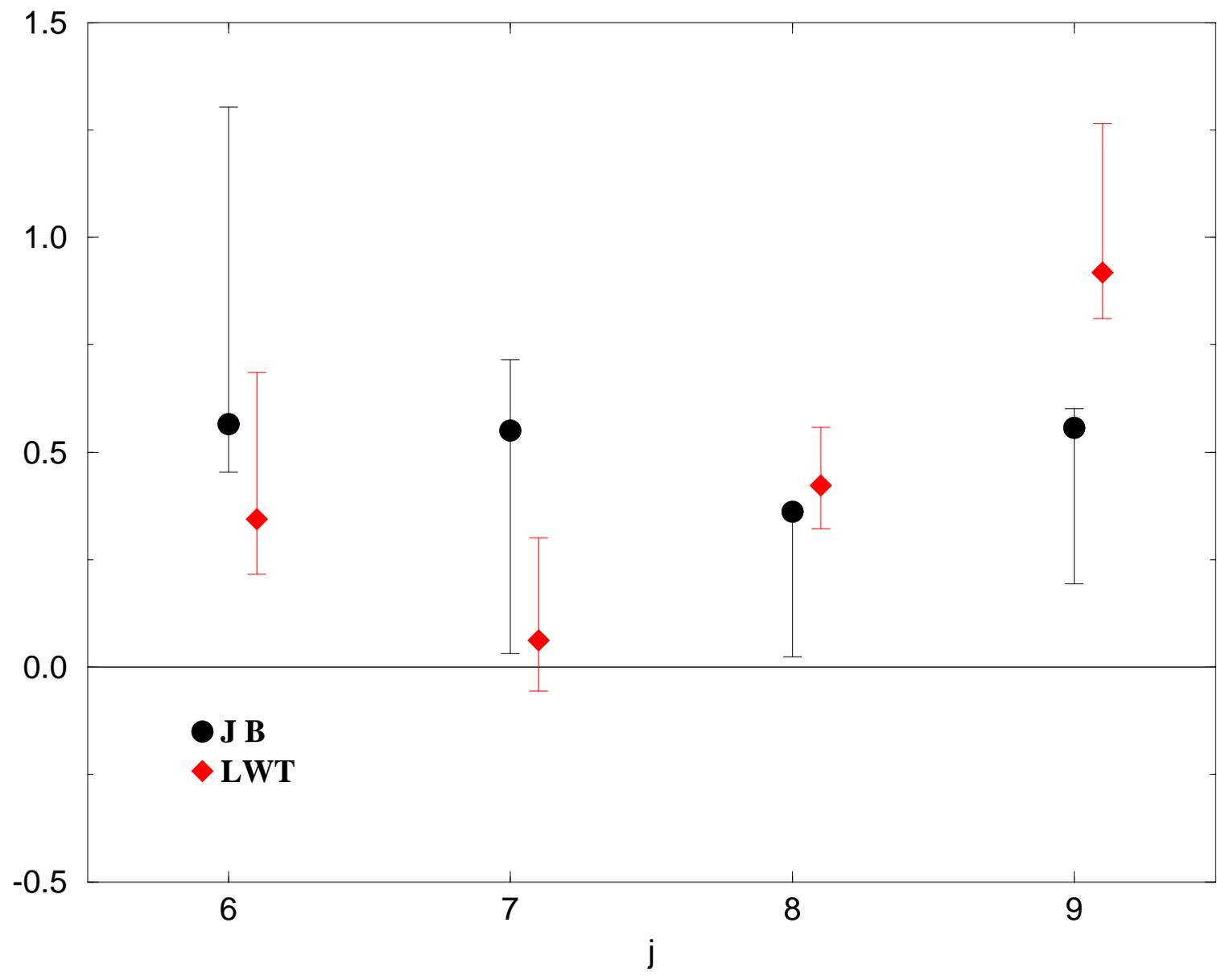
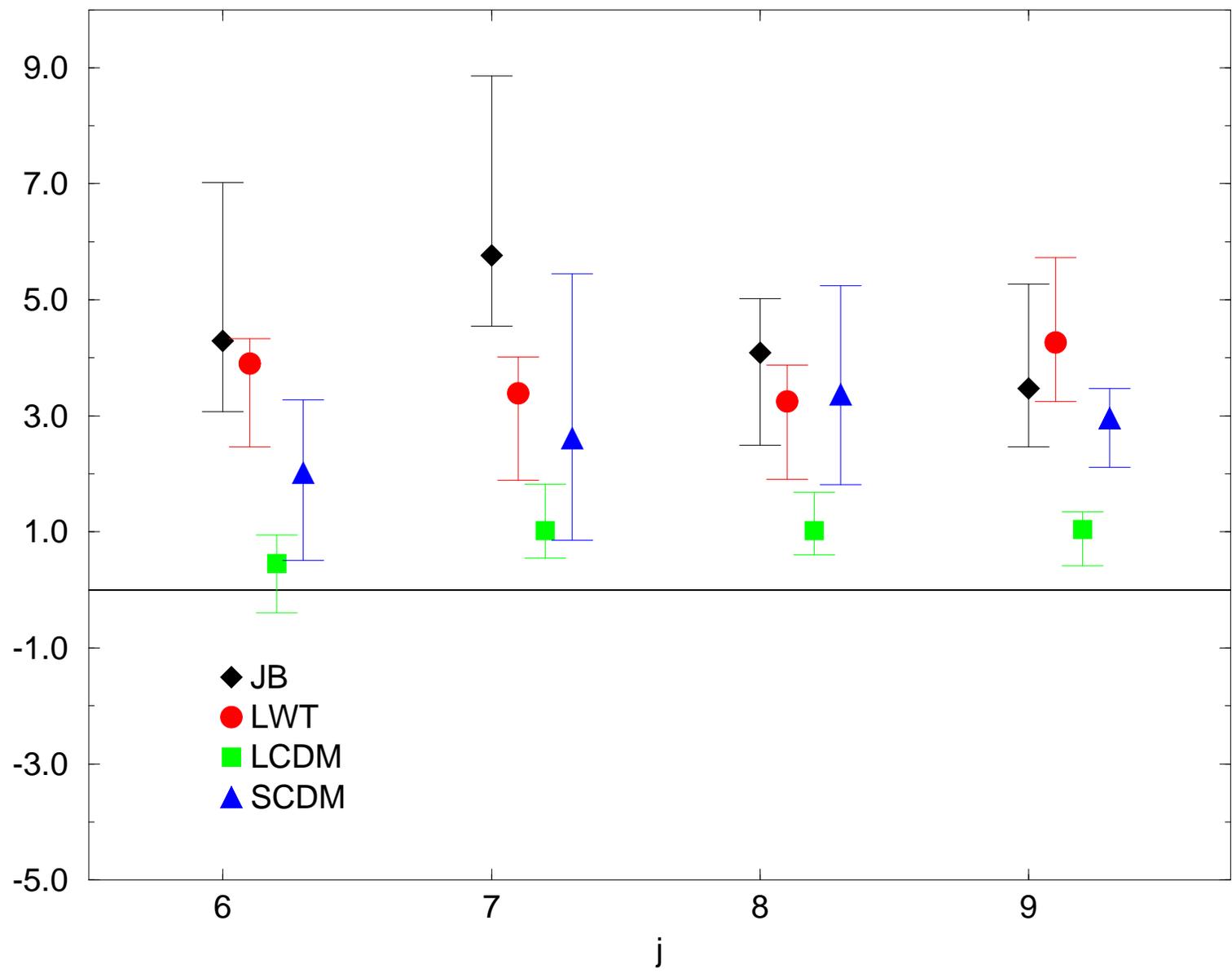
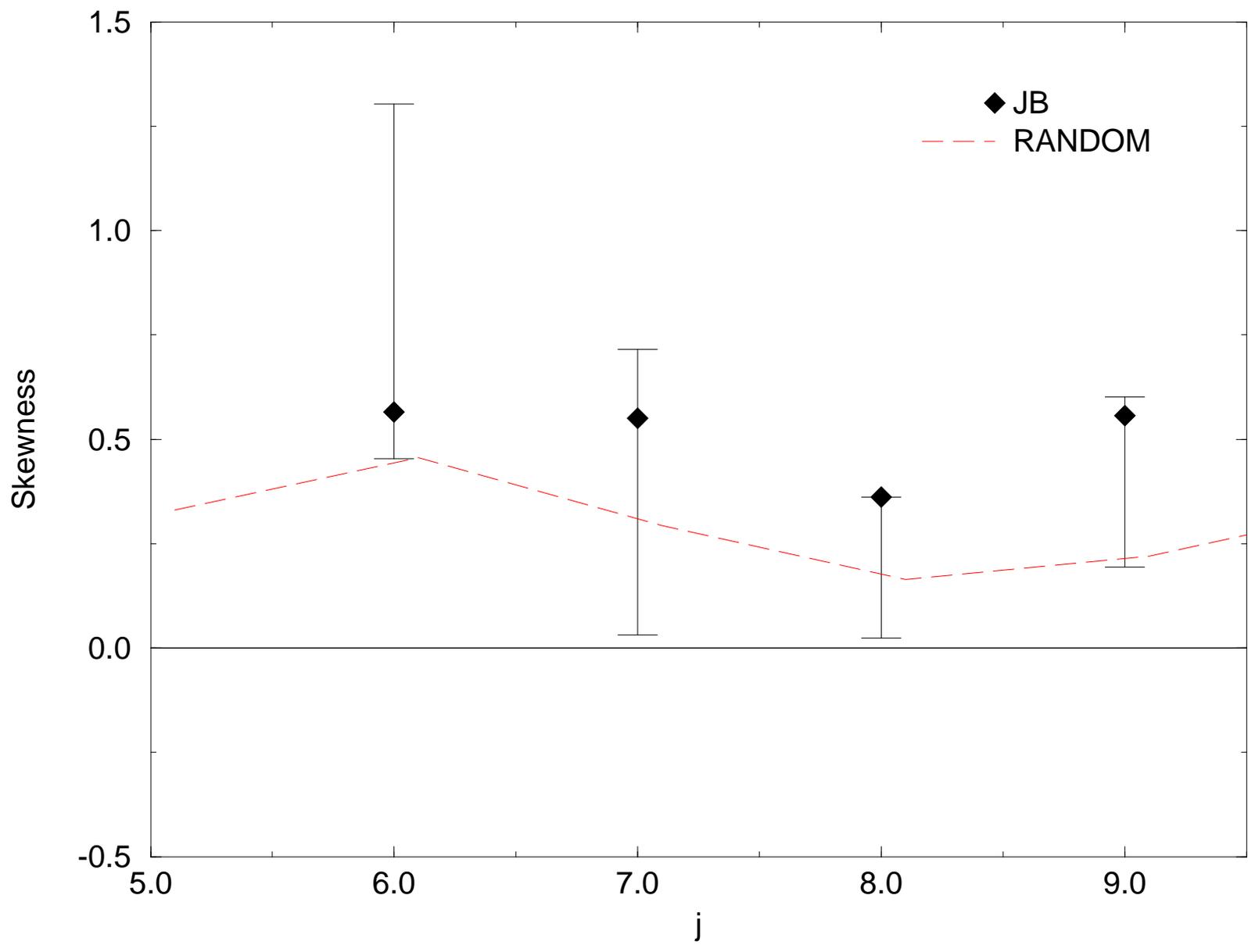
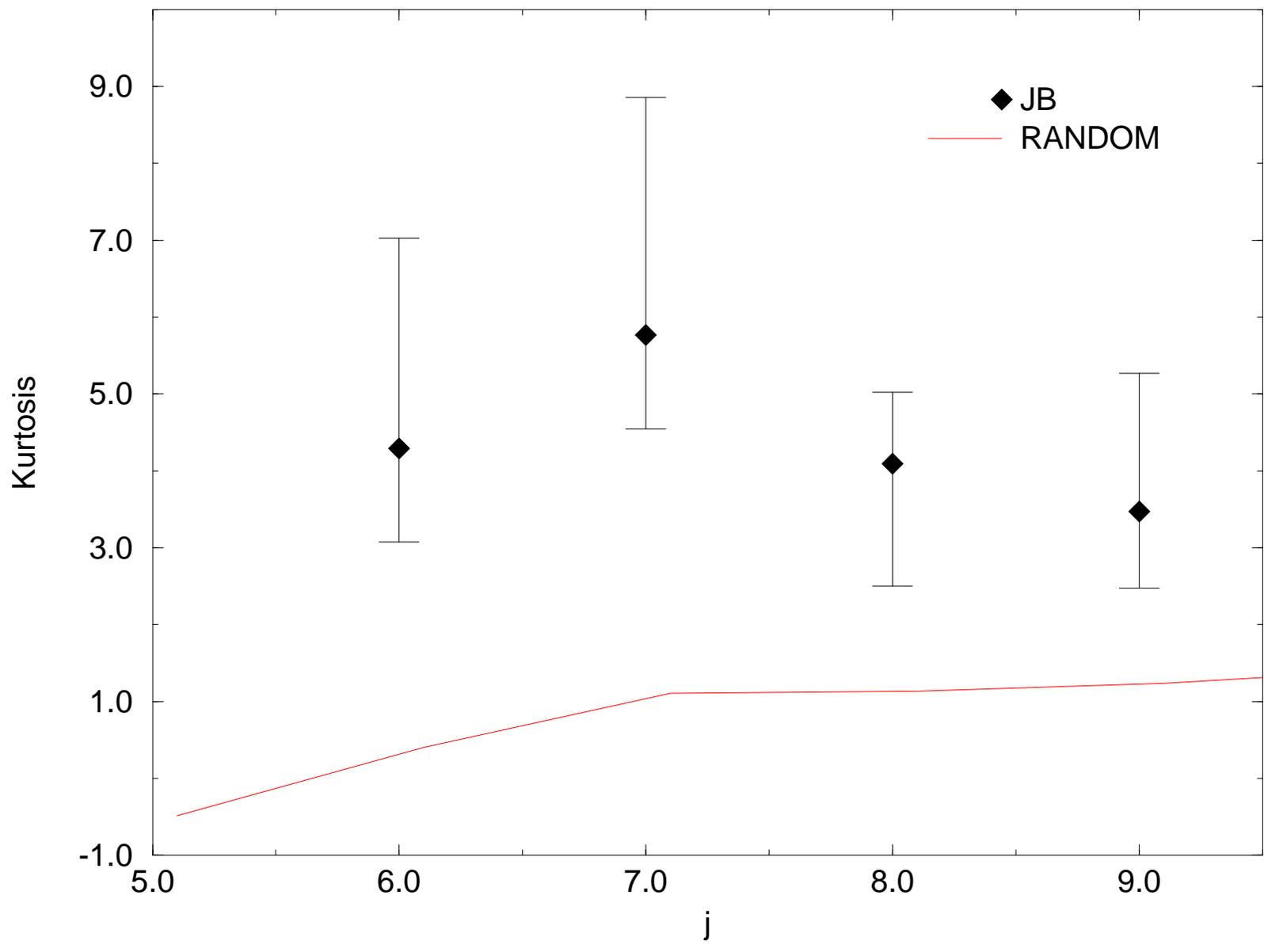
\end{document}